# A mathematical, *in silico* implemented, modular model for tumor growth in a spatially inhomogeneous, time-varying chemical environment.

1st Unrevised version.

Currently under revision.

Markos Antonopoulos[1] and Georgios Stamatakos[1]

[1]In Silico Oncology and In Silico Medicine Group, Laboratory of Microwaves and Fiber Optics, Institute of Communication and Computer Systems, School of Electrical and Computer Engineering, National Technical University of Athens, 9 Iroon Polytechniou, Zografou, 15780 Athens, Greece

Email addresses:

MA: markosan@central.ntua.gr

GS: gestam@central.ntua.gr

# Abstract

During the last decades, medical observations and multiscale data concerning tumor growth are mounting. At the same time, contemporary imaging techniques well established in clinical practice, provide a variety of information on real-time, *in-vivo* tumor growth. Mathematical and *in-silico* modeling has been widely recruited to provide means for further understanding of pertinent biological phenomena. However, despite the vast amounts of new evidence compiled by medical doctors, there are still many aspects of tumor growth that remain largely unknown. There is still a large variety of mechanisms to be better understood and therefore, many hypotheses to be tested. To approach this problem, starting from mathematical elaborations, we have developed a model of the early phases of tumor growth consisting of several algorithmic modules, each one corresponding to a particular biological mechanism. The modularity of the model allows keeping track of the assumptions made in each step and facilitates re-adjustment, in case new hypotheses need to be considered. Simulations showed good qualitative agreement with biological observations, and revealed a non-trivial interplay between between oxygen requirements of cancer cells and their maximum mitosis rates. The proposed model has, at least in principle, the potential to exploit data from contemporary imaging techniques and is eligible for utilizing multicore computation.

I.Introduction

Cancer is one of the main causes of mortality in the world. Statistics estimate that about one fifth of the population will suffer from cancer at some point of their lives [1]. Cancer is a category of diseases, which share several common features including sustained, uncontrolled cellular proliferation, resistance of cell death, induction of angiogenesis and activation of invasion and metastasis mechanisms [2].

The exact mechanisms that initiate cancer development remain largely unknown. However, it is widely accepted that cancer originates from cells which, due to various gene mutations, escape the body's natural mechanisms of controlling the balance between cell proliferation and cell death [3]. These cells create a clump which grows faster than host cells. However, this small tumor grows with a decreasing rate; as the tumor grows, disorganization of the host vasculature and limited diffusion of nutrients to the center of the tumor lead to the formation a necrotic core inside the tumor [4, 5]. Cells in the outer rim of the tumor proliferate, while cells in the interior die. For the tumor to grow up to becoming malignant, it needs to establish its own blood supply network, a process called angiogenesis. This process begins with tumor cells secreting molecules generally called tumor angiogenesis factors (TAF). TAFs induce proliferation and migration of endothelial cells, causing host vessels to form capillary sprouts. These sprouts randomly fuse with each other (anastomosis) and form loops, through which blood can flow. Repetition of these steps forms the tumor induced vasculature [6]. However, this tumor induced vasculature is highly disorganized, tortuous and dilated [7].

Evidently, during both the avascular and vascular phase of tumor development, the provision of nutrients to tumor cells through the blood supply network is highly inhomogeneous and time varying [4, 8, 9, 10]. In fact, many studies have shown that tumors contain hypoxic and hypoglycemic regions, particularly near the center which affect local cell proliferation and death rates [4, 11, 12 and references therein]. Contemporary imaging techniques such as positron emission tomography (PET) and magnetic resonance imaging (MRI) in its various forms (e.g. T1 gadolinium enhanced MRI, T2 MRI and Diffusion Weighted MRI) provide information on a variety of details concerning tumor microenvironment, such as glucose uptake rates and metabolism, oxygen distribution, cell density and proliferation. [13, 14, 15, 16]

During the last decades, in-silico modeling appears to be an invaluable tool for simulating complex biological processes, providing means for further understanding of related phenomena. In the following sections of this paper we develop a mathematical/ computational model that takes into account a variety of phenomena observed during the early (pre-neovascular) phases of tumor growth in a 3-dimensional, inhomogeneous and time-varying chemical environment. Since pertinent medical and biological literature on the subject shows that the vast majority of results are conclusive only up to a point, there are still a many qualitative and quantitative aspects of tumor progression that remain largely unknown, with angiogenesis-related phenomena forming a concrete class of such examples. Consequently, from a modeling perspective, it is almost impossible to avoid assumptions. Therefore, in this work, special care has been taken such that the model presented is modular, that is, it consists of several algorithmic modules, each one corresponding to a separate biological phenomenon encountered in tumor progression. The proposed methodology allows keeping track of the assumptions made in each part of the model, and facilitates model readjustment, in case new biological evidence emerge and need to be considered.

II. Related literature

Modelling the chemical microenvironment and growth of tumors has a long history. In the last 30 years, a plethora of models has appeared in the literature based on a variety of approaches, both continuous and discrete. In this section we review related literature. This review is by no means complete; it mainly describes previous work that inspired the work on this paper. For a comprehensive review, we refer to [17].

Early attempts in the subject include [18, 19]. In this work, the authors conducted a series of experiments and derived empirical equations quantifying the effects of oxygen and glucose concentrations and extracellular pH on cell metabolism and growth rate . Their mathematical approach was based on the reaction-diffusion equation, assuming spherical symmetry. They exploited the derived empirical equations by introducing appropriate terms taking account of cell metabolism, consumption of oxygen, aerobic/anaerobic breakdown of glucose and respective waste products. A theoretical study on the subject is that of [20]. In this work the authors used convection-diffusion equations and assumed spherical symmetry to model the spatial evolution of living and dead cells and how they are affected by the concentration of single generic nutrient. This approach was extended in [21, 22] which incorporated transport and metabolism of glucose, oxygen and lactate to their model. A different approach was taken in [23, 24]. In these works the authors used discrete, 2-dimensional cellular automata to account for single cell interactions with the environment and reaction-diffusion equations to model the evolution of chemical fields. In [25,26], the authors modeled cell surface receptors sensing the environment and subcellular molecular pathways by using an additional neural network determining the respective cellular phenotype. Spatially averaged cellular automata addressing additionally various treatment modalities include [27,28]. In a multiscale approach, the authors in [29] use a 3-dimensional lattice at the cellular level, a simplified protein regulatory network at the subcellular level and reaction-diffusion equations for the chemical fields. More recent similar approaches include [30,31,32]. Models explicitly incorporating vasculature and angiogenesis include [33,35].

Both continuum and discrete approaches were taken and compared in the series of papers [36,37]. In the continuum approach the authors used reaction-diffusion equations to model the spatiotemporal evolution of live and necrotic cells. In the discrete approach they used an off-lattice Voronoi/Delaunay cell model taking account of cell-cell interactions. Both approaches used reaction-diffusion equations to model the spatiotemporal evolution of glucose and oxygen concentrations. Interestingly, the authors conclude that when considering macroscopic quantities, continuous reaction-diffusion type models are adequate to explain global properties. According to the authors, agent-based models are difficult to discriminate, since most often, pertinent microscopic data regarding tissue morphology are not available. However, agent-based models have the ability to incorporate microscopic cellular properties determined by independently conducted experiments.

Our model is based on a systematic discretization of the continuous reaction- diffusion equations. We do not explicitly include vasculature and angiogenesis related phenomena in our model. As previously mentioned, angiogenesis is a highly complex phenomenon, and pertinent biological data are sparse. Instead, we model the effects of these phenomena on tumor growth by introducing macroscopic quantities like local nutrient provision, consumption and diffusion, which may be assessed or deduced by PET or MRI imaging techniques.

III. Modeling the diffusion of particles

A central notion, present in a vast variety of models regarding tumor growth, is that of diffusion. In this section, we will present a method to model this phenomenon, starting by the well-known reaction-diffusion partial differential equation.

$$\frac{\partial c}{\partial t} = \nabla \cdot (D \nabla c) \tag{1}$$

where $c(x,t)$ is the concentration of the particles under consideration (cells or molecules) at time $t$ and location $x$, and $D$ the diffusion tensor of the species in the surrounding material. This equation has been widely used to model cell diffusion, particularly in the case of glioblastoma, as well as diffusion of molecules in tissue.

In the case of isotropic diffusion D is a constant scalar, and equation (1) reduces to

$$\frac{\partial c}{\partial t} = D \Delta_x c \tag{2}$$

where $\Delta_x$ is the Laplace operator in $R^3$.

In [38] we elaborated on the observation that (2) is the Fokker-Planck equation corresponding to the stochastic differential equation

$$dx_t = \sqrt{2D} \cdot dB_t \tag{3}$$

where $B_t$ denotes standard Brownian motion in $R^3$. Given the initial position $x_o$ of a particle, the distribution of the random variable $x_t$ (i.e. the solution of (3) at time $t$) provides the probability distribution over all possible locations of this particle at time $t$. Given the initial position of a particle in terms of a probability distribution $c(x, 0)$, the probability distribution $c(x, t)$ can be found by two equivalent ways: By solving (2) as a partial differential equation with initial value $c(x, 0)$, to find the evolution of this distribution through time, or equivalently, by solving the stochastic differential equation (3) with initial distribution $c(x, 0)$ to find the probability distribution of the random variable $x_t$.

In case we want to study the movement of many particles, that is, molecules or cells located within a specified anatomic region, this notion of distribution is interpreted as follows: Integration of $c(x, t)$ over an area $A$ of $R^3$ yields the fraction of the total particle population that is located in $A$.

In the remaining part of this section, by furtherly elaborating on the previous observations, we will present a systematic discretization the above equations, in both time and space. This method will subsequently be used to simulate the evolution of phenomena that can be modeled by the preceding equations.

We start by considering the initial value problem

$$\frac{\partial c}{\partial t} = D \Delta_x c \tag{4}$$

$$c(x, t_o) = f(x) \tag{5}$$

We assume that

$$\int_{R^3} f(x) dx = 1 \tag{6}$$

i.e., f is a probability density function and that f(x) is supported in a compact subset $U \subseteq R^3$, i.e. $f(x) > 0$ if and only if $x \in U$. Equation (5) provides the initial value of the problem, that is, the distribution of the species in $R^3$ at $t = t_o$.

The problem defined by equations (4), (5) is well-posed and for $t \geq t_o$ has the solution

$$c(x,t) = \int_{R^3} G(x,y,t-t_o)f(y)dy = \int_U G(x,y,t-t_o)f(y)dy \tag{7}$$

Where $G(x,y,t-t_o)$ is the Gaussian kernel

$$G(x,y,t-t_o) = \frac{1}{\left(\sqrt{4\pi D(t-t_o)}\right)^3} e^{-\frac{\|x-y\|^2}{4D(t-t_o)}} \tag{8}$$

In probabilistic terms, equation (7) provides the probability density function (pdf) of the distribution of the species at time $t$ *given* that the probability density function of the respective distribution at $t = t_o$ is $f(y)$. Equivalently, we can calculate the probability for a particle to lie within a set $V \subseteq R^3$ (including the case $V \cap U \neq \emptyset$) at time t given that its position at $t = t_o$ is distributed according to f(y)

$Pr(particle\ lies\ in\ V\ at\ time\ t\ |\ particle's\ position\ initial\ pdf\ is\ f(y)) =$
$$\int_V \left( \int_U G(x,y,t-t_o)f(y)dy \right) dx \tag{9}$$

Each of the integrals that appear in equation (9) is a triple integral. The complete calculation implied in (9) can be carried out numerically.

We can drop the assumption in equation (6) and perform the same analysis. In this case, $f(x)$ represents the local density of particles at each point $x$. Integrating $f$ over $R^3$ provides the total population of the particles under consideration. The integral of $f$ over a subset $A \subseteq R^3$ provides the population of the particles that lies within $A$. Calculation of the integral

$$I = \int_V \left( \int_U G(x,y,t-t_o)f(y)dy \right) dx$$

yields the population that lies within the set $V$ at time $t$, when we know that the initial local density of the particles at time $t = t_o$ is $f$.

We proceed now to the discretization of the above equations. Let us assume that the diffusion of the particles under consideration takes place in a cubic lattice consisting of $N \times N \times N$ geometrical cells (voxels). Each voxel is a cube of dimensions $\Delta s \times \Delta s \times \Delta s$. We fix a temporal discretization step equal to $\Delta \tau$. Voxels in that cubic lattice can be classified into 4 categories, depending on the number of their neighboring voxels within the lattice. Voxels in the interior of the lattice have 26 neighbors. Voxels at the outer faces of the lattice have 17 neighbors. Voxels at the outer edges of the lattice have 11 neighbors and voxels at the outer vertices of the lattice have 7 neighbors. Additionally, for each particular voxel, its neighboring voxels fall into 3 categories: the ones that share a common face, the ones that share a common edge and the ones that share a common vertice with the particular voxel.

The first key step to the discretization process we propose is to assume that at each discrete time point, the distribution of the species under consideration within each voxel is uniform. Considering that the quantities of cells or molecules within a $mm^3$ are in the order of $10^6$ and $10^{12}$ respectively, we can deduce that this assumption is safe, as long the voxel size remains in the order of $mm^3$.

Under this assumption, at any discrete time point $t$ and for any pair of voxels $A$ and $B$ (not necessarily different) we can calculate the probability, for a particle to lie within $B$ at time $t + \Delta \tau$ *given* that it's position at time $t$ is a uniformly distributed (u.d.) random variable supported in $A$, by performing the calculation implied in equation (9) for $U = A, V = B$ and $f = (1/vol(A))1_A$. Here, $vol(A)$ is the volume of voxel $A$ ($= \Delta s^3$ for all voxels) and $1_A$ is the indicator function of $A$ as a subset of $R^3$.

Under the aforementioned assumption, equation (9) takes the form

$$Pr(particle\ lies\ in\ B\ at\ time\ t + \Delta\tau\ |\ particle\ lies\ in\ A\ at\ time\ t, u.d.) =$$
$$\int_B \left( \int_A G(x, y, \Delta\tau)(1/vol(A))1_A dy \right) dx \tag{10}$$

In the remaining part of the paper, for any pair of voxels A and B (not necessarily different) we will use the abbreviation $Pr(A \to B)$ for the probability implied in equation (10).

Supposing that the quantity of the particles within voxel $A$ at time $t$ is $Q$, the probability calculated from equation (10) can be interpreted as the fraction of $Q$ that will lie within voxel B at time $t + \Delta\tau$. Indeed, let us assume for the moment that $= (Q/vol(A))1_A$, that is, $f$ represents the (uniform) local density of particles within $A$ instead of a probability density function. As we mentioned previously, in this case, the integral $I$ yields the population that lies within the voxel B at time $t + \Delta\tau$, when we know that the local density of the species at time $t$ is $f$. We calculate

$$I = \int_V \left( \int_U G(x, y, \Delta\tau) f(y) dy \right) dx =$$
$$\int_B \left( \int_A G(x, y, \Delta\tau)(Q/vol(A))1_A dy \right) dx =$$
$$Q \cdot \int_B \left( \int_A G(x, y, \Delta\tau)(1/vol(A))1_A dy \right) dx =$$
$$Q \cdot Pr(A \to B)$$

where the last equality is inferred from equation (10).

Let $A$ denote a voxel not lying at the boundary of the lattice and $B_i$, $i = 1, ..., 26$ his neighbors. The second key step to our discretization process is to choose the voxels' edge length $\Delta s$ and the time step $\Delta\tau$ such that, from one time instant to the next one, for a voxel in the interior of the lattice, it's population diffuses at most in his neighbors. Mathematically, this means that $\Delta s$ and $\Delta\tau$ should be chosen such that

$$Pr(A \to A) + \sum_{i=1}^{26} Pr(A \to B_i) = 1 \tag{11}$$

Thus, for any voxel $A$ not lying at the boundary of the lattice, knowing the particle population of the voxel $Q_t(A)$ and its neighbors $Q_t(B_i)$, $i = 1, ..., 26$ at a time instant t, allows us calculate the population within A at the next time instant $t + \Delta\tau$ by

$$Q_{t+\Delta\tau}(A) = Pr(A \to A)Q_t(A) + \sum_{i=1}^{26} Pr(B_i \to A)Q_t(B_i) \tag{12}$$

For fixed $\Delta\tau$, in the discrete time framework, equation (12) is equivalently written

$$Q_{t_{p+1}}(A) = Pr(A \to A)Q_{t_p}(A) + \sum_{i=1}^{26} Pr(B_i \to A)Q_{t_p}(B_i) \tag{13}$$

Under the assumption that particles are uniformly distributed within each voxel and for properly chosen $\Delta s$ and $\Delta\tau$, equation (13) consists the discretization we propose and will be the basic tool we are going to use for modeling diffusion phenomena. The probabilities in this equation can be calculated numerically from the integral in (10). For remaining part of this section, we do not consider voxels lying at the boundary of the lattice nor voxels adjacent to it. We will deal with these voxels in detail in the next section, where we discuss boundary conditions.

Let the coordinates of each voxel in the lattice be given by a triad of integers, $(i, j, k)$ where $i, j, k = 1, ..., N$. We define a mapping $L: N^3 \to N$ as follows: $L(i, j, k) = i + (j - 1)N + (k - 1)N^2$. This mapping is a bijection from the space of triads $(i, j, k)$, $i, j, k = 1, ..., N$ to the integers from 1 to $N^3$. Let $Q(i, j, k)$ denote the number of particles in the respective voxel. We define a vector $q$ of $N^3$ elements by $q(L(i, j, k)) = Q(i, j, k)$, that is, we map the elements of the 3d matrix $Q$ to the one dimensional vector $q$.

Each voxel $A$ that is not lying at the boundary nor is adjacent to it, has 26 neighbors, which can be classified in 3 categories: 6 ones that share a common face with $A$, 12 that share a common edge with $A$ and 8 that share a common vertice with $A$. We denote these voxels by $F_l^A$, $l = 1, ..., 6$, $E_l^A$, $l = 1, ..., 12$, and $V_l^A$, $l = 1, ..., 8$, respectively. Due to symmetry and isotropy, (i.e. diffusion coefficient is a constant scalar) for any such voxel $A$ the following hold:

-For any $l_1, l_2 = 1, \ldots, 6$, $\Pr(F_{l_1}^A \to A) = \Pr(F_{l_2}^A \to A)$

-For any $l_1, l_2 = 1, \ldots, 12$, $\Pr(E_{l_1}^A \to A) = \Pr(E_{l_2}^A \to A)$

-For any $l_1, l_2 = 1, \ldots, 8$, $\Pr(V_{l_1}^A \to A) = \Pr(V_{l_2}^A \to A)$

We remind that each of these probabilities is defined and calculated by equations (8) and (10). Additionally, for any pair of voxels $A$ and $B$ not lying at the boundary nor adjacent to it, it holds

-$\Pr(A \to A) = \Pr(B \to B)$

-For any $l_1, l_2 = 1, \ldots, 6$, $\Pr(F_{l_1}^A \to A) = \Pr(F_{l_2}^B \to B)$

-For any $l_1, l_2 = 1, \ldots, 12$, $\Pr(E_{l_1}^A \to A) = \Pr(E_{l_2}^B \to B)$

-For any $l_1, l_2 = 1, \ldots, 8$, $\Pr(V_{l_1}^A \to A) = \Pr(V_{l_2}^B \to B)$.

The previous observations indicate that, given $q_p$, i.e. the vector of particle populations within each voxel at time instant $p$, by using equation (13) we can calculate each element of the vector $q_{p+1}$ that corresponds to a voxel not lying at the boundary, nor is adjacent to it, by using only four numbers: one for each neighbor category and one for the fraction of particles that were in $A$ and will remain in $A$. We denote these numbers by $\Pr(F \to A)$, $\Pr(E \to A)$, $\Pr(V \to A)$ and $\Pr(A \to A)$, respectively. These numbers can be precalculated by equations (8) and (10), where in each case, $D$ is taken to be the diffusion coefficient of the particles under consideration.

However, to be able to calculate the remaining elements of the vector $q_{p+1}$, i.e. those that correspond to voxels at the boundary or adjacent to it, one must consider boundary conditions. This is the subject of the following section.

IV. Dirichlet and Neumann boundary conditions.

Knowing the population of the particles within each voxel of the lattice at a time instant, equation (13) allows us to calculate the population within each voxel at the next time instant for all the voxels, except the ones at the boundary of the lattice and the ones that are adjacent to it. Each of the voxels lying at the boundary has less than 26 neighbors. In fact, each of these voxels would have 26 neighbors in an infinite lattice, but not all of them are included in a bounded, $N \times N \times N$ lattice. Unless we specify boundary conditions, the calculation in (13) cannot be performed neither for those voxels and consequently, nor their neighbors. There are two types of boundary conditions that can be imposed in a classical diffusion problem, those of the Dirichlet type and those of the Neumann type.

Dirichlet boundary conditions express the requirement that at the boundary of the region under consideration, the quantity of interest does not change with time. In the framework presented here, this is expressed mathematically by the following: If a voxel $A$ lies at the boundary of the lattice, to calculate its population at the next time instant, instead of equation (13) simply apply $Q_A(t_{p+1}) = Q_A(t_p)$. Additionally, if a voxel is adjacent to the boundary, simply apply (13) by using the respective probabilities as they are calculated from (8) and (10).

Neumann boundary conditions express the requirement that at the boundary of the region under consideration, the flux of the quantity of interest is zero, i.e. the boundary is non-permeable. In terms of calculus this is expressed by the requirement, at any time instant and at any point of the boundary, the projection of the gradient of the quantity on the outward normal of the boundary at that point to be zero. In the stochastics literature, a non-permeable boundary within which a random motion takes place is often referred as a reflecting boundary [39]. In the framework presented here, this is expressed mathematically as follows.

As previously mentioned, any voxel $A$ lying at the boundary has 17, 11 or 7 neighbor voxels which we denote by $B_i$, where $i$ is an integer from 1 to 17,11 or 7, depending on the position of the voxel. For each such voxel $A$, we calculate the probabilities appearing in equation (13) only for the voxels that are contained in the lattice. Specifically, if $A$ lies at the boundary, we calculate the probabilities $\Pr(A \to A)$, $\Pr(A \to B_i)$ where $i$ is an integer from 17, 11 or 7. We then normalize these probabilities to sum to one. By this calculation we acquire the probabilities we need, in order to apply equation (13) for any voxel in the lattice. Using these normalized probabilities when applying equation (13) for either boundary voxels, or their neighbors, ensures that every particle lying in a voxel at the boundary, will remain within the lattice, that is, within the region of interest, corresponding to the notion of reflecting boundary.

In our case, the region of interest is a cube; It is apparent, that these methods of imposing Dirichlet or Neumann boundary conditions apply to more complex shapes, as long as they are properly discretized. The case of glioblastoma, where the scull naturally imposes a reflecting boundary to the diffusion of glioma cells is an example where this approach may be useful.

V. Diffusion of Glucose and Oxygen.

In this section we will use the previously developed ideas to model the diffusion of chemical molecules in the region of interest, that is, the cubic lattice of dimensions $N \times N \times N$. Assuming Dirichlet boundary conditions, equation (13) and the observations in section II imply that if we know $q_p$, i.e. the quantity of the molecules within each voxel at a time instant $t_p$, we can calculate $q_{p+1}$ by performing a linear calculation. This means that there is a $N^3 \times N^3$ square matrix $T$ such that $q_{p+1} = T q_p$.

We remind that for each voxel $A$ not at the boundary, we can apply equation (13) by using only four numbers, which we respectively denote by $\Pr(F \to A)$, $\Pr(E \to A)$, $\Pr(V \to A)$ and $\Pr(A \to A)$. These numbers can be precalculated by equations (8) and (10), where in each case, $D$ is taken to be the diffusion coefficient of the respective molecule. We subsequently use these values to construct the matrix $T$ according to the following algorithm:

Algorithm 1

$T = N^3 \times N^3$ zero matrix           //initialization
**for** each triad $(i,j,k)$  $i,j,k = 1, \ldots, N$
  **if** the voxel with coordinates $(i,j,k)$ is at the boundary
  $T(L(i,j,k), L(i,j,k)) = 1$           // Dirichlet boundary condition
  **else**
    **for** $di = -1,0,1$  $dj = -1,0,-1$  $dk = -1,0,1$
      **if**    $|di| + |dj| + |dk| = 0$
        $T(L(i,j,k), L(i,j,k)) = \Pr(A \to A)$
      **elseif** $|di| + |dj| + |dk| = 1$           //common face neighbor
        $T(L(i,j,k), L(i+di, j+dj, k+dk)) = \Pr(F \to A)$
      **elseif** $|di| + |dj| + |dk| = 2$           //common edge neighbor
        $T(L(i,j,k), L(i+di, j+dj, k+dk)) = \Pr(E \to A)$
      **elseif** $|di| + |dj| + |dk| = 3$           //common vertice neighbor
        $T(L(i,j,k), L(i+di, j+dj, k+dk)) = \Pr(V \to A)$
      **end**
    **end**
  **end**
**end**

To summarize, for each chemical species, we use (8) and (10) to calculate the four probabilities $\Pr(F \to A)$, $\Pr(E \to A)$, $\Pr(V \to A)$ and $\Pr(A \to A)$. We then apply the previous algorithm, to construct the respective matrix. In our model, we will use two such matrices, one for glucose and one for oxygen, denoted by $T_{gl}$, $T_o$ respectively. We will use the notations $gl_p$, $o_p$ to denote the each of $N^3$ vectors, whose elements are the molar quantities of glucose and oxygen within each voxel at time instant $t_p$.

For each case, the matrix constructed by the previous algorithm has some interesting properties. First, by equation (11) and the properties of the probabilities $\Pr(F \to A)$, $\Pr(E \to A)$, $\Pr(V \to A)$ and $\Pr(A \to A)$ presented in the section I, it is a stochastic matrix, i.e. the sum of each row is 1. Second, for sufficiently large $N$, and due to the assumption implied in equation (11), it is a sparse matrix: in each row, at most 26 elements are nonzero. This will provide a significant relief in the computational burden of the entire model.

## VI. The diffusion of living cells

In the model we propose, there are two types of cell populations within each voxel, the living cells and the necrotic cells. In view of the mapping $L$ and the notation established in section III, let $l_p$, $n_p$ denote the $N^3 \times 1$ vectors, whose entries are the populations of living and necrotic cells within each voxel at time instant $t_p$. Let $u_p$ denote the sum of $l_p$ and $n_p$, i.e. the total (live+necrotic) cell population within each voxel.

Starting from equation (13), we will construct an algorithm for simulating the diffusion of live cells. We do not directly apply the method developed in the previous section for two reasons. The first reason is that we assume that only living cells are able to move. The second reason is that we assume that for any movement of cells between two neighboring voxels to happen, the total number of cancer cells (live +necrotic) should have reached a critical population in at least one of the two voxels. The implementation of a somewhat different hypothesis is presented in the appendix. In what follows, we denote that critical population by $C$. Special care has been taken such that each term of the equations presented below, has a natural and intuitive meaning.

First we remind that, due to symmetry and the fact that we assume isotropic diffusion (i.e. $D$ is a constant scalar) for any pair of neighboring voxels $A$, $B$ that do not lie at the boundary of the lattice nor adjacent to it, it holds $Pr(A \to B) = Pr(B \to A)$. For that particular case, we adopt the notation

$$\Pr(A \to B) = \Pr(B \to A) = \Pr(A \leftrightarrow B) = \Pr(B \leftrightarrow A) \tag{14}$$

Taking into account equations (11), (14), for any voxel A not lying at the boundary of the lattice nor is adjacent to it, equation (13) takes the equivalent form

$$Q_A(t_{p+1}) = Q_A(t_p) + \sum_{i=1}^{26} \Pr(A \leftrightarrow B_i)(Q_{B_i}(t_p) - Q_A(t_p)) \tag{15}$$

Verbally, equation (15) states that, the difference between cell populations at two consecutive time instants within a voxel, is the algebraic sum of the net numbers of cells that moved between the voxel and each of its neighbors. If the voxel contains more cells than a particular neighbor, cells move from the voxel to the neighbor. Conversely, if the voxel contains fewer cells than a particular neighbor, cells move from the neighbor to the voxel. Note that based on the observations in section III, to apply equation (15) for voxels not lying in the boundary of the lattice nor are adjacent to it, we only need four numbers, which again, can be precalculated numerically from equations (8) and (10).

Let as now assume that we know $l, n, u$ at time instant $t_p$. Let $next\_l$ denote the vector of living cell populations due to diffusion at the next time instant. We use the notation $l(A)$, $n(A)$, $u(A)$ to denote the living, dead and total cell population within a voxel $A$. The vector of live cell populations due to diffusion of live cells at the next time instant is calculated by the following algorithm:

Algorithm 2

**for** each voxel $A$ in the lattice
  **if** voxel $A$ and each of its neighbors have zero total population
    $next\_l(A) = l(A)$                // no diffusion
  **else**
    $s = l(A)$
    **for** each of voxel's $A$ neighbors $B_i$
      **if** $u(A) \geq C$ or $u(B_i) \geq C$
        **if** $u(B_i) \geq u(A)$
$$s = s + \frac{l(B_i)}{u(B_i)} \Pr(A \leftrightarrow B_i)(u(B_i) - u(A))$$
        **else**
$$s = s + \frac{l(A)}{u(A)} \Pr(A \leftrightarrow B_i)(u(B_i) - u(A))$$
        **end**
      **end**
    **end**
    $next\_l(A) = s$
  **end**
**end**
$l = next\_l$

The algorithm checks each voxel. If the voxel and its neighbors are all of zero total population, no movement of cells happens. If not, all pairs $(A, B_i)$ of the voxel and its neighbors are processed sequentially. For any movement of cells between the voxel and a particular neighbor to happen, at least one of them must have total cell population above the critical. If the voxel has lower total population than the particular neighbor, the quantity $\Pr(A \leftrightarrow B_i)(u(B_i) - u(A))$ provides the net number of cells that would have moved from $B_i$ to $A$, if all cells within $B_i$ were alive: However, this is not always the case. Therefore we multiply this quantity by the ratio $l(B_i)/u(B_i)$. If the voxel has higher total population than the particular neighbor, the quantity $\Pr(A \leftrightarrow B_i)(u(B_i) - u(A))$ provides the net number of cells that would have moved from $A$ to $B_i$, if all cells within $A$ were alive. Again, since this is not always the case, we multiply this quantity by $l(A)/u(A)$. Note that the total population of living cells before and after the algorithm does not change: what is added to voxel $A$ from $B_i$, will be substracted from $B_i$ when the algorithm checks voxel $B_i$ and its neighbors. What is substracted from voxel $A$ to move to $B_i$, will be added to $B_i$ when the algorithm checks voxel $B_i$ and its neighbors.

Essentially, the algorithm is based on equation (15). However, for any particular voxel and any of its neighbors, the algorithm checks additional conditions before adding the respective term to the algebraic sum in (15).

We note that by choosing a sufficiently large lattice and placing the initial tumor in the center of it, there is no need to impose Neumann conditions in the sense of section 2. In that case, cancer cells do not reach the boundary voxels nor their neighbors and for all time steps these voxels are always processed by the first if-statement of the algorithm. Thus, we can safely apply the previously described method. However, imposing Neumann boundary conditions in the sense of section 2 is totally feasible. The respective algorithmic implementation is more involved. To avoid interrupting the flow and keep the presentation simple, we provide the details in the appendix.

Both algorithm 2 and algorithm 6 in the appendix are built on a spatially distributed dynamical systems perspective. Each element of the next state vector is calculated using only the elements of the previous state vector. Thus, each such calculation can be performed independently of the others rendering algorithms 2 and 6 suitable for exploiting multicore computation.

V. Proliferation/necrosis of cancer cells and oxygen/glucose consumption

In this section, we build the part of the model which at each time step, for each separate voxel, calculates the proliferation and necrosis of cancer cells, their consumption of glucose and oxygen, and their respective quantities within the voxel at the next time instant, neglecting for the moment diffusion-related phenomena. We will include diffusion-related calculations in the following section, where we present the complete structure the model. At each time step $p$, the input of this part consists of the four $N^3 \times 1$ vectors $l_p, n_p, gl_p, o_p$ and the output of the respective vectors $l_{p+1}, n_{p+1}, gl_{p+1}, o_{p+1}$. Based on the literature, we will make several assumptions and build an algorithm that distinguishes a variety of cases depending on cancer cell populations and oxygen/ glucose levels in the voxel under consideration.

We introduce the following parameters:
- $M$, the average cell population capacity (cancerous +normal) per voxel.
- $K_{gl}$, the average consumption of glucose for a normal cell of the host non-proliferating tissue. (units: pmol/sec)
- $K_o$, the average consumption of oxygen for a normal cell of the host non-proliferating tissue. (units: pmol/sec)
- $\lambda_1, \lambda_2$, coefficients assuming positive values, such that $\lambda_1 K_{gl}$, $\lambda_2 K_o$ are the average consumptions of glucose and oxygen for an actively proliferating cancer cell. Non-proliferating cancer cells are assumed to consume the same amounts of glucose and oxygen as the host tissue cells.
- $a_{max}$ the generic, prefixed maximum mitosis rate of cancer cells per time step $\Delta \tau$. This parameter reflects the maximum mitosis rate of the cancer cells, when glucose and oxygen levels are ideal.

We will assume that initially, each voxel contains $M$ normal cells. As the tumor grows, these cells become either dislocated or apoptotic. Each normal cell consumes $K_{gl}$ pmol glucose and $K_o$ pmol oxygen per second. This means that initially, in the absence of cancer cells, the blood supply network should provide in average $MK_{gl}$ and $MK_o$ pmols of glucose and oxygen per second in each voxel, in order to keep oxygen and glucose levels at a constant level [40,41]. Let $F_{gl}, F_o$ be the initial quantity of glucose and oxygen within each voxel. Mathematically, this means that at time $p = 0$, for each voxel $A$ in the lattice, $gl_0(A) = F_{gl}$ and $o_0(A)=F_o$.

The stoichiometry of the clean combustion of glucose requires that the glucose/oxygen uptake ratio is 1:6. It is well documented that for cancer cells, due to increased utilization of glycolysis, this is not the case [2]. Experimental measurements and estimations report that the ratio glucose /oxygen consumption in tumors can vary up to 1:1 or even more [18, 28, 29, 36, 42, 44]. Compared to clean combustion, glycolysis is 18 times less efficient in ATP production and cancer cells compensate this deficiency by upregulating glucose transporters, thereby increasing glucose import in the cytoplasm. Considerably increased glucose uptake and utilization has been reported in a variety of tumors by the use of positron emission tomography (PET) [2]. It is also reported that local levels of oxygen and glucose have an effect on this ratio [18, 43]. Quantitative details of this effect are still unclear. The possibility that a single cell may employ both glycolysis and normal aerobic metabolism is not excluded. Qualitatively it seems evident, that when oxygen falls below a certain threshold, cells tend to switch to a glycolytic phenotype. However, this observation does not tell the whole story, since cancer cells switch to glycolysis even when oxygen levels are abundant [45,46]. In the cellular automaton model proposed in [27,28] the authors assumed that each cell can employ one of the two metabolic pathways, and when the local oxygen level is zero cells switch from aerobic metabolism to glycolysis. We opt for a different approach, more similar to [29]; as described in detail in what follows, for each simulation we fix the parameters $\lambda_1$ and $\lambda_2$ as average values and use them to calculate the proliferation of cells within each voxel as they are confined by the local levels of glucose and oxygen. It will become apparent that this may be easily readjusted; it is totally feasible to introduce an additional function determining the *local* $\lambda_1$ and $\lambda_2$ for the cells of a voxel by the voxel's glucose and oxygen levels. We leave this for future work. Following several authors cited in the literature, we will assume that when oxygen or glucose levels within a voxel fall below certain respective thresholds, cells within that voxel start becoming necrotic.

As previously mentioned, tumor vasculature is highly irregular, and relative medical data are sparse. However, qualitative observations indicate that tumor growth affects preexisting vasculature even before the onset of angiogenesis. Vessels incorporated in the tumor mass may disintegrate or get obstructed [4]. The precise qualitative and quantitative details of this phenomenon remain largely unknown. However, since the focus in this work is on macroscopic quantities like local nutrient provision from the local blood supply network, a quantification of these effects on local host vasculature, and consequently on local nutrient provision, should be included. It is observed that for several tumors, the concentration of oxygen in the necrotic core of tumor varies between 0.5% and 30% of the concentration in the surrounding tissue [5,27,28]. Thus, at a first approximation, we will make the assumption that if a voxel contains a percentage of necrotic tissue, i.e. a percentage of its *M* cells is necrotic, the provision of glucose and oxygen through the blood supply network is reduced by the same percentage. We note that this assumption represents a correlational, rather than a causational relation between these quantities. Although this approach is somewhat oversimplifying, the entire structure of the algorithm enables readjustment so as to apply for more sophisticated scenarios, e.g. taking into account total cell density within a voxel or random fluctuations in the voxels' local nutrient provision due to disorganized blood supply.

Let us know suppose that at time instant $p$ we know $l_p, n_p, gl_p, o_p$. Let $\Delta\tau$ be the time step of the simulation. For each voxel $A$ in the lattice, we distinguish the following cases.

<u>Case 1</u>. $l_p(A) = 0, n_p(A) = 0$. In this case, the quantities of glucose and oxygen consumed by the normal cells within $A$ are provided by the blood supply network. Glucose and oxygen levels within $A$ do not change during $\Delta\tau$. Thus,

$$l_{p+1}(A) = l_p(A)$$
$$n_{p+1}(A) = n_p(A)$$
$$gl_{p+1}(A) = gl_p(A)$$
$$o_{p+1}(A) = o_p(A)$$

<u>Case 2</u>. Glucose or oxygen levels are below certain respective thresholds. These states are known as hypoglycemia and hypoxia, respectively. It has been observed that in hypoxic areas within tumors the oxygen levels vary from 0.5% to 30% of the normal oxygen levels in non-tumor areas. A typical respective threshold for hypoglycemia is 50% [47]. We will denote by $h_o, h_{gl}$ these percentages. Thus, if $gl_p(A) < h_{gl}F_{gl}$ ($A$ is hypoglycemic) or $o_p(A) < h_o F_o$ ($A$ is hypoxic) living cancer cells should stop proliferating and start becoming necrotic. Cells that stay alive consume glucose and oxygen provided by the blood supply and so, their respective levels remain constant. Let $r_n$ be the necrosis rate. We have

$$l_{p+1}(A) = l_p(A) - r_n l_p(A)$$
$$n_{p+1}(A) = n_p(A) + r_n l_p(A)$$
$$gl_{p+1}(A) = gl_p(A)$$
$$o_{p+1}(A) = o_p(A)$$

Case 3. $l_p(A) > 0, n_p(A) = 0$.

In this case we have several subcases:

Case 3.1. $l_p(A) \leq M$, that is, living cancer cells within $A$ are below the average cell capacity of the voxel. Let $a$ be the mitosis rate of living cancer cells per time step $\Delta\tau$, i.e. the fraction of cells within $A$ that will divide during $\Delta\tau$ and $cc$ the duration of their cell cycle. The duration of their cell cycle in time steps is $cc/\Delta\tau$. Assuming a uniform distribution of proliferating cells at all phases of the cell cycle, we estimate a total number $a \cdot l_p(A) \cdot (cc/\Delta\tau)$ of actively proliferating cancer cells. Since $M$ is the average cell capacity of the voxels, apart from cancer cells, $A$ contains $M - l_p(A)$ normal cells. We calculate the consumption rates of glucose and oxygen by all cells in $A$.

- $G = MK_{gl} + a \cdot l_p(A) \cdot (cc/\Delta\tau)(\lambda_1 - 1)K_{gl}$ pmol glucose/sec. The quantity $MK_{gl}$ is the amount of glucose consumed by cells in $A$ per second that is provided by the blood supply. The quantity

$$G_1 = a \cdot l_p(A) \cdot (cc/\Delta\tau)(\lambda_1 - 1)K_{gl}$$

is the amount of glucose consumed by cells in $A$ per second that is not provided by the blood supply. The quantity $\Delta\tau \cdot G_1$ is the respective amount of glucose consumed by cells in $A$ during the entire time step $\Delta\tau$. For any proliferation to happen, cancer cells should be able to acquire this amount from the quantity of glucose within $A$, i.e. $gl_p(A)$

- $O = MK_o + a \cdot l_p(A) \cdot (cc/\Delta\tau)(\lambda_2 - 1)K_o$ pmol oxygen/sec. The quantity $MK_o$ is the amount of oxygen consumed by cells in $A$ per second that is provided by the blood supply. The quantity

$$O_1 = a \cdot l_p(A) \cdot (cc/\Delta\tau)(\lambda_2 - 1)K_o$$

is the amount of oxygen consumed by cells in $A$ per second that is not provided by the blood supply. The quantity $\Delta\tau \cdot O_1$ is the respective amount of oxygen consumed by cells in $A$ during the entire time step $\Delta\tau$. For any proliferation to happen, cancer cells should be able to acquire this amount from the quantity of oxygen within $A$, i.e. $o_p(A)$.

Subsequently, we solve the inequalities

$$gl_p(A) - \Delta\tau \cdot G_1 \geq 0.05 \cdot F_{gl} \quad \text{and} \quad o_p(A) - \Delta\tau \cdot O_1 \geq 0.05 \cdot F_o$$

with respect to $a$; Let $a \leq a_{gl}$, $a \leq a_o$ be the solutions. The common solutions of these inequalities provide the maximum mitosis rate that $a$ that cancer cells can achieve at these levels $gl_p(A)$, $o_p(A)$ of glucose and oxygen. We set $a = min(a_{gl}, a_o)$. We note that the right-hand sides of these inequalities could have been zero. To avoid negative quantities that may appear from rounding during the simulation, we have chosen to insert these quantities as small (0.05, may be chosen to be even lower) percentages of the concentrations of glucose and oxygen in normal tissue. If the maximum mitosis rate $a$ is larger than $a_{max}$, we set $a = a_{max}$, else $a$ remains as calculated from the above inequalities, i.e. $a = min(a_{gl}, a_o)$. Finally, for the next time instant $p+1$, we calculate

$$l_{p+1}(A) = l_p(A) + a l_p(A)$$
$$n_{p+1}(A) = n_p(A)$$
$$gl_{p+1}(A) = gl_p(A) - \Delta\tau \cdot G_1$$

$$o_{p+1}(A) = o_p(A) - \Delta\tau \cdot O_1$$

We note that Case 3 should be always checked after Case 2. This ensures that the maximum mitosis rates $a_{gl}, a_o$ as calculated from the previous inequalities are nonnegative.

<u>Case 3.2</u> $l_p(A) > M$, i.e. living cancer cells within $A$ are above the average cell capacity of the voxel. Using the notation established in Case 3.1, we first calculate the consumption rates of glucose and oxygen by all cells in $A$.

- $G = MK_{gl} + (l_p(A) - M)K_{gl} + a \cdot l_p(A) \cdot (cc/\Delta\tau)(\lambda_1 - 1)K_{gl}$ pmol glucose/sec. The quantity $MK_{gl}$ is the amount of glucose consumed by cells in $A$ per second that is provided by the blood supply. The quantity
$$G_1 = (l_p(A) - M)K_{gl} + al_p(A)(cc/\Delta\tau)(\lambda_1 - 1)K_{gl}$$
is the amount of glucose consumed by cells in $A$ per second that is not provided by the blood supply. The quantity $\Delta\tau \cdot G_1$ is the respective amount of glucose consumed by cells in $A$ during the entire time step $\Delta\tau$. For any proliferation to happen, cancer cells should be able to acquire this amount from the quantity of glucose within $A$, i.e. $gl_p(A)$

- $O = MK_o + (l_p(A) - M)K_o + a \cdot l_p(A) \cdot (cc/\Delta\tau)(\lambda_2 - 1)K_o$ pmol oxygen/sec. The quantity $MK_o$ is the amount of oxygen consumed by cells in $A$ per second that is provided by the blood supply. The quantity
$$O_1 = (l_p(A) - M)K_o + a \cdot l_p(A) \cdot (cc/\Delta\tau)(\lambda_2 - 1)K_o$$
is the amount of oxygen consumed by cells in $A$ per second that is not provided by the blood supply. The quantity $\Delta\tau \cdot O_1$ is the respective amount of oxygen consumed by cells in $A$ during the entire time step $\Delta\tau$. For any proliferation to happen, cancer cells should be able to acquire this amount from the quantity of oxygen within $A$, i.e. $o_p(A)$.

Again, we solve the inequalities
$$gl_p(A) - \Delta\tau \cdot G_1 \geq 0.05 \cdot F_{gl} \quad \text{and} \quad o_p(A) - \Delta\tau \cdot O_1 \geq 0.05 \cdot F_o$$
with respect to $a$; Let $a \leq a_{gl}, a \leq a_o$ be the solutions. There is no guarantee that $a_{gl}, a_o$ are nonnegative; Therefore, we distinguish two further subcases.

<u>Case 3.2.1</u> Both $a_{gl}, a_o$ are nonnegative. In this case, we set $a = min(a_{gl}, a_o)$. If the maximum mitosis rate $a$ is larger than $a_{max}$, we set $a = a_{max}$, If not, it remains $a = min(a_{gl}, a_o)$.

For the next time instant $p+1$, we calculate
$$l_{p+1}(A) = l_p(A) + al_p(A)$$
$$n_{p+1}(A) = n_p(A)$$
$$gl_{p+1}(A) = gl_p(A) - \Delta\tau \cdot G_1$$
$$o_{p+1}(A) = o_p(A) - \Delta\tau \cdot O_1$$

<u>Case 3.2.2</u> At least one of $a_{gl}, a_o$ is negative. In view of the previous inequalities resulting in $a_{gl}, a_o$, this implies that at least one of the quantities
$$gl_p(A) - 0.05 \cdot F_{gl} - \Delta\tau \cdot (l_p(A) - M)K_{gl}$$
$$o_p(A) - 0.05 \cdot F_o - \Delta\tau \cdot (l_p(A) - M)K_o$$
is negative. This means that cells within $A$ need more resources (oxygen, glucose or both) than those available $(gl_p(A), o_p(A))$ just for staying alive, without proliferation. We proceed as follows. First, we note that from the fact that we do not have necrotic tissue in the voxel under consideration i.e. $n_p(A) = 0$, blood supply is adequate for all $M$ of $l_p(A)$ cells to survive $(l_p(A) \geq M)$. We divide $gl_p(A)/(\Delta\tau \cdot K_{gl})$ and $o_p(A)/(\Delta\tau \cdot K_o)$ and set $n_s = min(gl_p(A)/(\Delta\tau \cdot K_{gl}), o_p(A)/(\Delta\tau \cdot K_o))$. The number $n_s$ is the number of cells additional to $M$ that can survive during $\Delta\tau$ at these at these levels $gl_p(A)$, $o_p(A)$ of glucose and oxygen. Again, to avoid negative quantities that may appear from rounding during the simulation, we set $n_s = 0.9 \cdot n_s$. The coefficient 0.9 may be chosen to be larger, as long as it is smaller than 1. Thus, apart from the $M$ living cells, an additional number of $n_s$ cells survives with no proliferation, consuming the necessary amounts of glucose and oxygen. Mathematically, for the next time instant $p+1$, this means.

$$l_{p+1}(A) = M + n_s$$
$$n_{p+1}(A) = (l_p(A) - M) - n_s$$
$$gl_{p+1}(A) = gl_p(A) - n_s \cdot \Delta\tau \cdot K_{gl}$$
$$o_{p+1}(A) = o_p(A) - n_s \cdot \Delta\tau \cdot K_o$$

<u>Case 4</u> $l_p(A) \geq 0, n_p(A) > 0$.
We have several subcases.
<u>Case 4.1</u> $n_p(A) \geq M$, i.e. the necrotic population within $A$ is larger than the average cell capacity of the voxel.
<u>Case 4.1.1</u> $n_p(A) \geq M$ and $l_p(A) \geq 0$. In this case, $A$ contains more necrotic cells than its average cell capacity. We consider the voxel necrotic and that there is no blood reaching it. We calculate the quantities $gl_p(A)/(\Delta\tau \cdot K_{gl})$, $o_p(A)/(\Delta\tau \cdot K_o)$ and set $n_s = \min(gl_p(A)/(\Delta\tau \cdot K_{gl}), o_p(A)/(\Delta\tau \cdot K_o))$. $n_s$ is the number of cells that can survive during $\Delta\tau$ at these at these levels $gl_p(A)$, $o_p(A)$ of glucose and oxygen. Again, to avoid negative quantities that may appear from rounding during the simulation, we set $n_s = 0.9 \cdot n_s$. There are two special subcases.
<u>Case 4.1.1.1</u> $l_p(A) \leq n_s$. In this case, the number of cells that can survive during $\Delta\tau$ is larger than the living cell population within $A$. Thus, all living cells within $A$ survive with no proliferation, consuming the necessary amounts of glucose and oxygen. Mathematically, this is expressed by the calculation

$$l_{p+1}(A) = l_p(A)$$
$$n_{p+1}(A) = n_p(A)$$
$$gl_{p+1}(A) = gl_p(A) - n_s \cdot \Delta\tau \cdot K_{gl}$$
$$o_{p+1}(A) = o_p(A) - n_s \cdot \Delta\tau \cdot K_o$$

<u>Case 4.1.1.2</u> $l_p(A) > n_s$. In this case, glucose and oxygen levels do not suffice to keep all living cells alive during $\Delta\tau$. Only $n_s$ of $l_p(A)$ cells will survive, consuming the necessary amounts of glucose and oxygen. For the next time instant $p+1$, we calculate

$$l_{p+1}(A) = n_s$$
$$n_{p+1}(A) = n_p(A) + (l_p(A) - n_s)$$
$$gl_{p+1}(A) = gl_p(A) - n_s \cdot \Delta\tau \cdot K_{gl}$$
$$o_{p+1}(A) = o_p(A) - n_s \cdot \Delta\tau \cdot K_o$$

<u>Case 4.1.2</u> $n_p(A) \geq M$ and $l_p(A) = 0$. In this case, voxel $A$ contains exclusively necrotic cells. Again, we consider the voxel necrotic and that there is no blood reaching it. For the next time instant $p+1$, everything remains as it is.

$$l_{p+1}(A) = l_p(A)$$
$$n_{p+1}(A) = n_p(A)$$
$$gl_{p+1}(A) = gl_p(A)$$
$$o_{p+1}(A) = o_p(A)$$

<u>Case 4.2</u> $n_p(A) < M$. There are the following special subcases.

<u>Case 4.2.1</u> $n_p(A) < M$ and $l_p(A) \leq M$. In this case, $A$ contains a nonzero number of necrotic cells. We will assume that in this case, the amount of glucose and oxygen that reaches $A$ is $(M - n_p(A))/M$ times the amount in the case of zero necrosis within $A$. As in the cases 3.1 and 3.2, we calculate the consumption rates of glucose and oxygen by all cells in $A$.

- $G = (M - n_p(A))K_{gl} + a \cdot l_p(A) \cdot (cc/\Delta\tau)(\lambda_1 - 1)K_{gl}$ pmol glucose/sec. The quantity $(M - n_p(A))K_{gl}$ is the amount of glucose consumed by cells in $A$ per second that is provided by the blood supply.
  The quantity
  $$G_1 = a \cdot l_p(A) \cdot (cc/\Delta\tau)(\lambda_1 - 1)K_{gl}$$
  is the amount of glucose consumed by cells in $A$ per second that is not provided by the blood supply. The quantity $\Delta\tau \cdot G_1$ is the respective amount of glucose consumed by cells in $A$ during the entire time step $\Delta\tau$. For any proliferation to happen, cancer cells should be able to acquire this amount from the quantity of glucose within $A$, i.e. $gl_p(A)$

- $O = (M - n_p(A))K_o + a \cdot l_p(A) \cdot (cc/\Delta\tau)(\lambda_2 - 1)K_o$ pmol oxygen/sec. The quantity $(M - n_p(A))K_o$ is the amount of oxygen consumed by cells in $A$ per second that is provided by the blood supply.
  The quantity
  $$O_1 = a \cdot l_p(A) \cdot (cc/\Delta\tau)(\lambda_2 - 1)K_o$$
  is the amount of oxygen consumed by cells in $A$ per second that is not provided by the blood supply. The quantity $\Delta\tau \cdot O_1$ is the respective amount of oxygen consumed by cells in $A$ during the entire time step $\Delta\tau$. For any proliferation to happen, cancer cells should be able to acquire this amount from the quantity of oxygen within $A$, i.e. $o_p(A)$.

As in the cases 3.1 and 3.2, we subsequently solve the inequalities
$$gl_p(A) - \Delta\tau \cdot G_1 \geq 0.05 \cdot F_{gl} \text{ and } o_p(A) - \Delta\tau \cdot O_1 \geq 0.05 \cdot F_o$$
with respect to $a$; Let $a \leq a_{gl}$, $a \leq a_o$ be the solutions. We set $a = min(a_{gl}, a_o)$. This is the maximum mitosis rate that cancer cells can achieve at these levels $gl_p(A)$, $o_p(A)$ of glucose and oxygen. If the maximum mitosis rate $a$ is larger than $a_{max}$, we set $a = a_{max}$, else $a$ remains as calculated from the above inequalities, i.e. $a = min(a_{gl}, a_o)$. For the next time instant $p+1$, we calculate
$$l_{p+1}(A) = l_p(A) + al_p(A)$$
$$n_{p+1}(A) = n_p(A)$$
$$gl_{p+1}(A) = gl_p(A) - \Delta\tau \cdot G_1$$
$$o_{p+1}(A) = o_p(A) - \Delta\tau \cdot O_1$$
We note that Case 4.2.1 should be always checked after Case 2. This ensures that mitosis rates $a_{gl}, a_o$ as calculated from the previous inequalities are nonnegative.

<u>Case 4.2.2</u> $n_p(A) < M$ and $l_p(A) > M$, i.e. the voxel contains a nonzero number of necrotic cells and a population of living cells larger than the average cell capacity of the voxel. We will again assume that the amount of glucose and oxygen that reaches $A$ is $(M - n_p(A))/M$ times the amount in the case of zero necrosis within $A$. We calculate the consumption rates of glucose and oxygen by all cells in $A$.

- $G = MK_{gl} + (l_p(A) - M)K_{gl} + a \cdot l_p(A) \cdot (cc/\Delta\tau)(\lambda_1 - 1)K_{gl} =$
  $= \left(M - n_p(A)\right)K_{gl} + n_p(A)K_{gl} + (l_p(A) - M)K_{gl} +$
  $+ a \cdot l_p(A) \cdot (cc/\Delta\tau)(\lambda_1 - 1)K_{gl}$ pmol glucose/sec.

  The quantity $(M - n_p(A))K_{gl}$ is the amount of glucose consumed by cells in $A$ per second that is provided by the blood supply. The quantity

$$G_1 = n_p(A)K_{gl} + (l_p(A) - M)K_{gl} + a \cdot l_p(A) \cdot (cc/\Delta\tau)(\lambda_1 - 1)K_{gl}$$
is the amount of glucose consumed by cells in $A$ per second that is not provided by the blood supply. The quantity $\Delta\tau \cdot G_1$ is the respective amount of glucose consumed by cells in $A$ during the entire time step $\Delta\tau$. For any proliferation to happen, cancer cells should be able to acquire this amount from the quantity of glucose within $A$, i.e. $gl_p(A)$

- $O = MK_o + (l_p(A) - M)K_o + a \cdot l_p(A) \cdot (cc/\Delta\tau)(\lambda_2 - 1)K_o =$
  $= \big(M - n_p(A)\big)K_o + n_p(A)K_o + (l_p(A) - M)K_o +$
  $+ a \cdot l_p(A) \cdot (cc/\Delta\tau)(\lambda_1 - 1)K_o$ pmol oxygen/sec.

The quantity $(M - n_p(A))K_o$ is the amount of oxygen consumed by cells in $A$ per second that is provided by the blood supply.

The quantity
$$O_1 = n_p(A)K_o + (l_p(A) - M)K_o + a \cdot l_p(A) \cdot (cc/\Delta\tau)(\lambda_2 - 1)K_o$$
is the amount of oxygen consumed by cells in $A$ per second that is not provided by the blood supply. The quantity $\Delta\tau \cdot O_1$ is the respective amount of oxygen consumed by cells in $A$ during the entire time step $\Delta\tau$. For any proliferation to happen, cancer cells should be able to acquire this amount from the quantity of oxygen within $A$, i.e. $o_p(A)$.

We subsequently solve the inequalities
$$gl_p(A) - \Delta\tau \cdot G_1 \geq 0.05 \cdot F_{gl} \quad \text{and} \quad o_p(A) - \Delta\tau \cdot O_1 \geq 0.05 \cdot F_o$$
with respect to $a$; Let $a \leq a_{gl}, a \leq a_o$ be the solutions. There is no guarantee that $a_{gl}, a_o$ are nonnegative; Therefore, we distinguish two further subcases.

<u>Case 4.2.2.1</u> Both $a_{gl}, a_o$ are nonnegative. In this case, we set $a = min(a_{gl}, a_o)$. If the maximum mitosis rate $a$ is larger than $a_{max}$, we set $a = a_{max}$, If not, it remains $a = min(a_{gl}, a_o)$.

For the next time instant $p+1$, we calculate
$$l_{p+1}(A) = l_p(A) + al_p(A)$$
$$n_{p+1}(A) = n_p(A)$$
$$gl_{p+1}(A) = gl_p(A) - \Delta\tau \cdot G_1$$
$$o_{p+1}(A) = o_p(A) - \Delta\tau \cdot O_1$$

<u>Case 4.2.2.2</u> At least one of $a_{gl}, a_o$ is negative. In view of the previous inequalities resulting in $a_{gl}, a_o$, this implies that at least one of the quantities
$$gl_p(A) - 0.05 \cdot F_{gl} - \Delta\tau \cdot \big(n_p(A)K_{gl} + (l_p(A) - M)K_{gl}\big)$$
$$o_p(A) - 0.05 \cdot F_o - \Delta\tau \cdot \big(n_p(A)K_o + (l_p(A) - M)K_o\big)$$
is negative. As in case 3.2.2, this means that cells within $A$ need more resources (oxygen, glucose or both) than those available ($gl_p(A)$, $o_p(A)$) just for staying alive, without proliferation. We note that blood supply is adequate for all $M - n_p(A)$ of $l_p(A)$ cells to survive ($l_p(A) \geq M \geq M - n_p(A)$). We divide $gl_p(A)/(\Delta\tau \cdot K_{gl})$ and $o_p(A)/(\Delta\tau \cdot K_o)$ and set $n_s = min(gl_p(A)/(\Delta\tau \cdot K_{gl}), o_p(A)/(\Delta\tau \cdot K_o))$. The number $n_s$ is the number of cells additional to $M - n_p(A)$ that can survive during $\Delta\tau$ at these at these levels $gl_p(A)$, $o_p(A)$ of glucose and oxygen. Again, to avoid negative quantities that may appear from rounding during the simulation, we set $n_s = 0.9 \cdot n_s$. Thus, apart from the $M - n_p(A)$ living cells, an additional number of $n_s$ cells survives with no proliferation, consuming the necessary amounts of glucose and oxygen. Mathematically, for the next time instant $p+1$, this means
$$l_{p+1}(A) = \big(M - n_p(A)\big) + n_s$$

$$n_{p+1}(A) = n_p(A) + \left(l_p(A) - \left(\left(M - n_p(A)\right) + n_s\right)\right)$$
$$gl_{p+1}(A) = gl_p(A) - n_s \cdot \Delta\tau \cdot K_{gl}$$
$$o_{p+1}(A) = o_p(A) - n_s \cdot \Delta\tau \cdot K_o$$

All the above cases are checked and the respective calculations are performed at each time instant, for each voxel in the lattice. We note that each separate case considered could be readjusted, in case different hypotheses need to be taken into account. This is made clear in the following section.

VI. The complete model

The model we propose can be seen as a discrete time dynamical system. The state of the system consists of the four $N^3 \times 1$ vectors $l_p, n_p, gl_p, o_p$. In sections III, IV, V we have defined four operators.

- The operator defined algorithmically in section V, which we denote by $O_1$. Applying operator $O_1$ to the state vector $(l_p, n_p, gl_p, o_p)$ at each time instant, consists in checking all cases described in section V and performing the respective calculations for any voxel in the lattice.
- The operators defined in section III, which we denote by $O_2, O_3$. These are actually the matrices $T_{gl}, T_o$ defined in section III. Applying one of these operators to the state vector $(l_p, n_p, gl_p, o_p)$ consists in multiplying the respective matrix ($T_{gl}$ or $T_o$) with the respective vector $gl_p$ or $o_p$, thereby calculating how the respective molecule quantities within each voxel change due to diffusion.
- The operator defined algorithmically in section IV, which we denote by $O_4$. Applying operator $O_4$ to the state vector $(l_p, n_p, gl_p, o_p)$ consists in executing Algorithm 2 (or Algorithm 6 from the Appendix) for the given vectors $l_p, n_p$, thereby calculating how living cell populations within each voxel change due to cell diffusion.

Knowledge of these vectors allows us to calculate $l_{p+1}, n_{p+1}, gl_{p+1}, o_{p+1}$ by the repeated application of the algorithm described in the following Diagram.

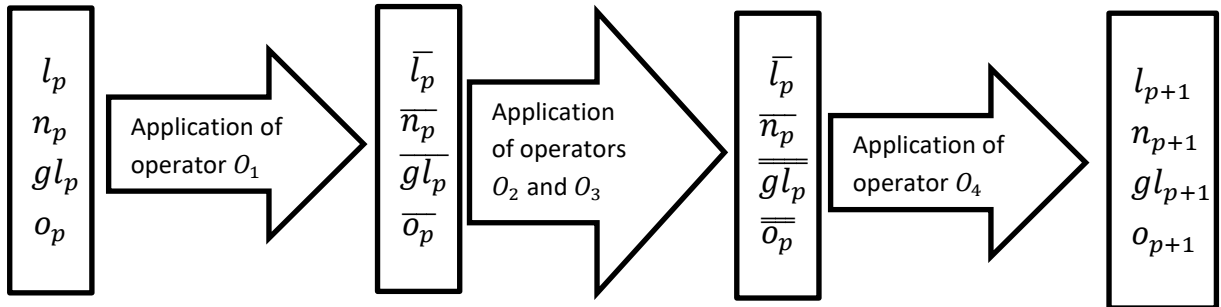

The entire architecture of the model is in compliance with approach presented in [48] and consists an extensible framework for modeling tumor growth. Phenomena pertaining to diffusion of cells and molecules are modeled by separate operators (i.e. algorithmic modules) applied sequentially to the state vector. We note that, at least in principle, the application of these operators to the state vector is parallelizable. Within the proposed methodology, the algorithmic module corresponding to operator $O_1$ is completely re-adjustable. This facilitates the simulation of scenarios based on different hypotheses concerning cell proliferation, necrosis and metabolism of chemical species. Introduction of additional diffusion operators like $O_2, O_3$ and extension of operator $O_1$ according to different assumptions enables consideration of additional chemical species such as lactate and growth factors. Introduction of additional cellular species is also feasible, by appropriate readjustment of operators $O_4$ and $O_1$.

As noted in [23], there is a large disparity between the time scales of cell proliferation and diffusion of chemical species. Instead of solving the reaction diffusion equations for each chemical species explicitly in time, the authors in [23] approached this problem by solving a sequence of equilibrium diffusion equations in a coarser time-scale. The method we present here, benefits from the sparse matrices constructed in section II and for small $\Delta\tau$ (in the order of 5-10 sec) affords a tractable computational implementation of both chemical fields evolution and cell proliferation. The problem is that for $\Delta\tau$ in the order of 5-10 sec, applying $O_4$ in each time step inflicts a significant computational burden in the entire simulation. However, diffusion coefficient of glucose and oxygen is in the order of $10^{-5}$ cm$^2$/sec while the respective coefficient for cancer cells is in the order of $10^{-8}$ cm$^2$/sec, reflecting a similar disparity in the time scales of chemical species and cell diffusion [21,27,28,36,37,49]. This allows applying operator $O_4$ every several time steps, corresponding to a respective time scale of minutes, thereby substantially improving computational time. In the simulations described in the next section, we used a 10 sec time step, and applied $O_4$ every 300 time-steps, i.e. 5 mins.

VI. Simulations

In this section we will use the model developed previously to simulate various scenarios, primarily concerning the needs of proliferating cancer cells in glucose and oxygen relative to the normal quiescent cells and how they affect tumor growth. Quantitatively, it has been observed that proliferating cells consume 2-5 more resources than quiescent cells [30, 29, 43]. Therefore, we will vary the parameters $\lambda_1$ and $\lambda_2$ from in the interval from 1 to 6. Time step and voxel edge where chosen $\Delta\tau = 10$ sec and $\Delta s = 2$ mm. respectively. These values satisfy the assumptions in section I. The time step could be chosen to be even lower (down to 1-2 secs), with some extra, although not severe impact on overall computational time. Furthermore, a voxel edge of 2 mm corresponds to the spatial resolution of contemporary MRI techniques [50]. The parameter values are presented in table 1. All simulations where initialized with a cancer cell population of $4\cdot10^6$ cells.

Table 1

| Parameter | Symbol | Value | References and remarks |
|---|---|---|---|
| Lattice size | $N$ | 21 | |
| Time step | $\Delta\tau$ | 10 sec | |
| Voxel edge | $\Delta s$ | 2 mm | |
| Average cell population capacity per voxel | $M$ | $8\cdot10^6$ cells/voxel | Dionysiou et al 2004 |
| Oxygen Diffusion Coefficient | $D_o$ | $1.8\cdot10^{-5}$ cm$^2$/sec | Gerlee&Anderson 2007,2008, Venkatasubramanian et al 2006, Schaller & Meyer-Hermann 2005, 2006 |
| Glucose Diffusion Coefficient | $D_{gl}$ | $1.05\cdot10^{-5}$ cm$^2$/sec | Gerlee&Anderson 2007,2008, Venkatasubramanian et al 2006, Schaller & Meyer-Hermann 2005, 2006 |
| Cell diffusion Coefficient | $D_c$ | $1.5\cdot10^{-8}$ to $1.5\cdot10^{-7}$ cm$^2$/sec | Murray 2001 |
| Quiescent host cell Oxygen consumption | $K_o$ | $50\cdot10^{-6}$ pmol/sec | Kempf et al 2015, Freyer & Sutherland 1985, Wherle et 2000 |
| Quiescent host cell Glucose consumption | $K_{gl}$ | $130\cdot10^{-6}$ pmol/sec | Kempf et al 2015, Freyer & Sutherland 1985, Wherle et 2000 |
| Critical Population | $C$ | $8\cdot10^6$ cells | Dionysiou et al 2004 |
| Cell cycle duration | $cc$ | 24 hours | Gerlee&Anderson 2007,2008 |
| Hypoxia threshold percentage | $h_o$ | 0.15 | Gerlee&Anderson 2007,2008 |

| Hypoglycemia threshold percentage | $h_{gl}$ | 0.40 | Gerlee&Anderson 2007,2008 |
| --- | --- | --- | --- |
| Maximum mitosis rate | $a_{max}$ | $8 \cdot 10^{-6}$ to $16 \cdot 10^{-6}$ mitoses/10 sec | Corresponds to a doubling time of 5 to 10 days in ideal chemical conditions |
| Necrosis rate | $r_n$ | $1.6 \cdot 10^{-5}$ cell deaths/ 10 sec | Corresponds to a 5 day half-life of dying cells |
| Oxygen per voxel in normal tissue | $F_o$ | $1.2 \cdot 10^3$ pmol | Kempf et al 2015, corresponding to concentration in capillary blood. |
| Glucose per voxel in normal tissue | $F_{gl}$ | $40 \cdot 10^3$ pmol | Kempf et al 2015, corresponding to concentration in capillary blood. |

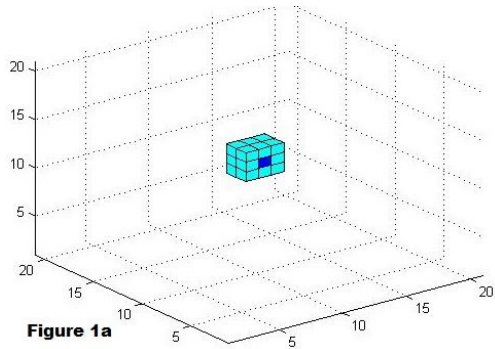
Figure 1a
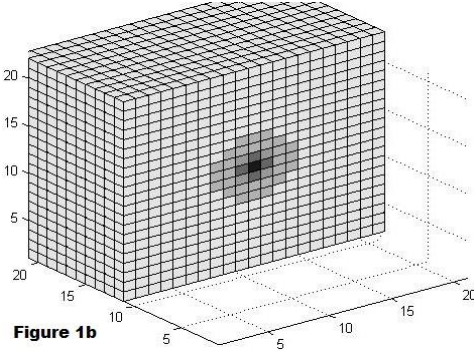
Figure 1b
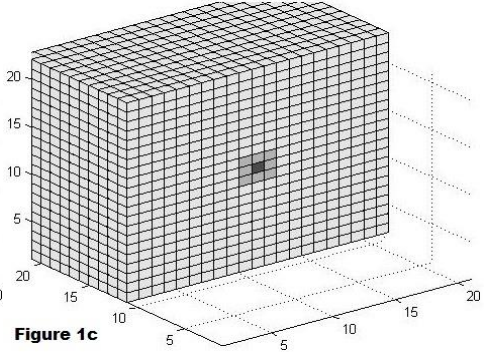
Figure 1c

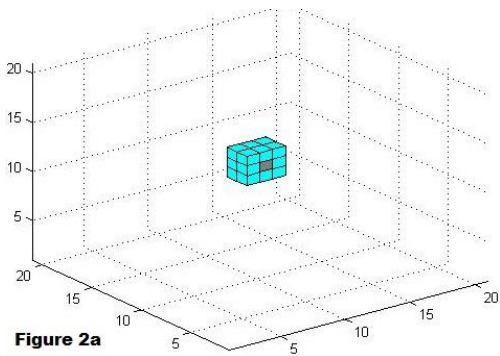
Figure 2a
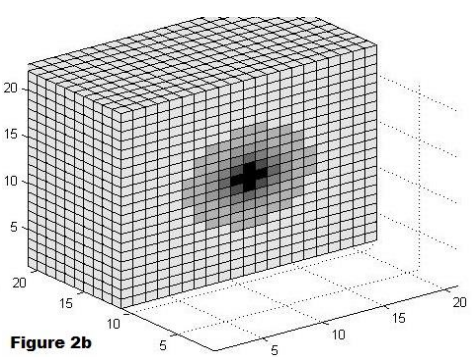
Figure 2b
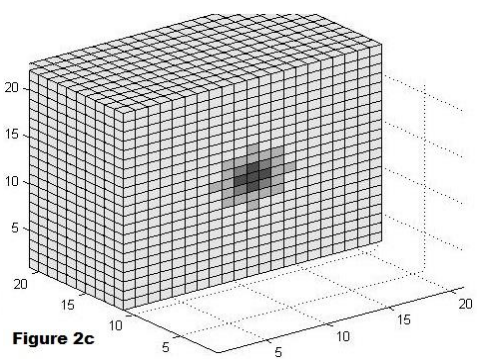
Figure 2c

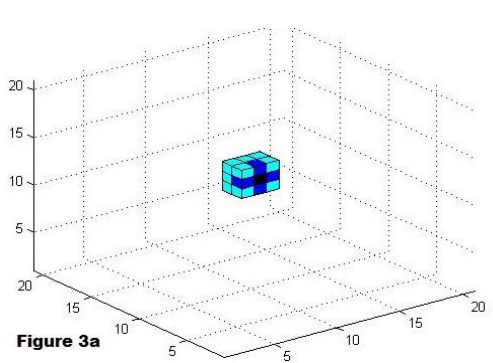
Figure 3a
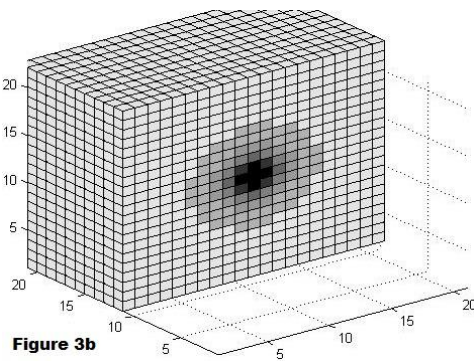
Figure 3b
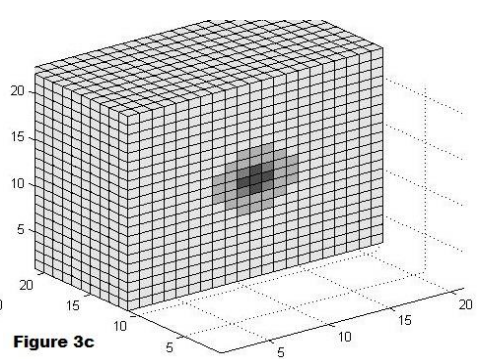
Figure 3c

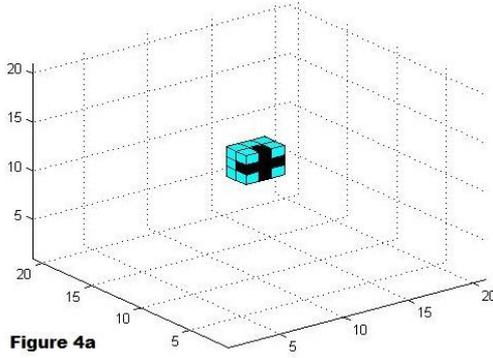 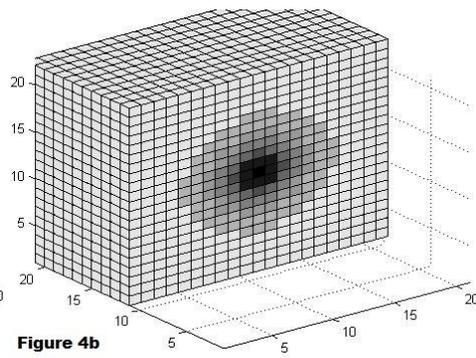 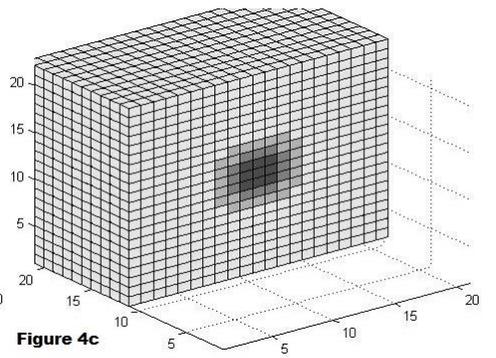
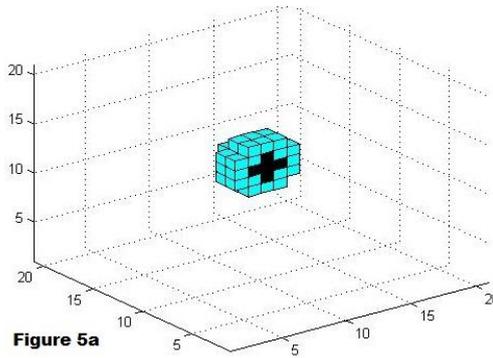 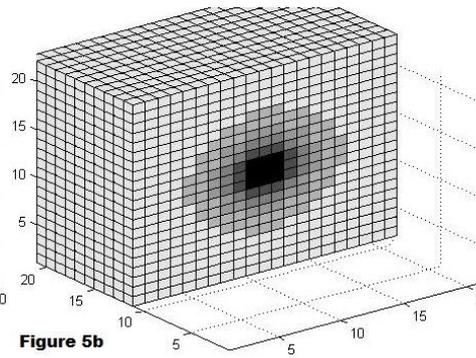 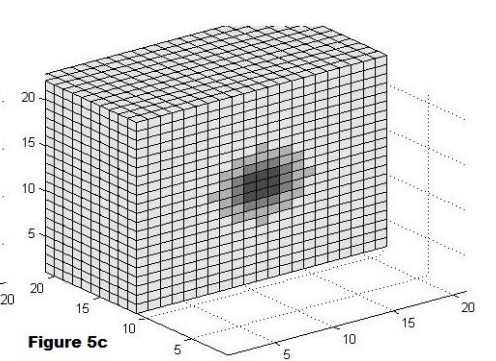
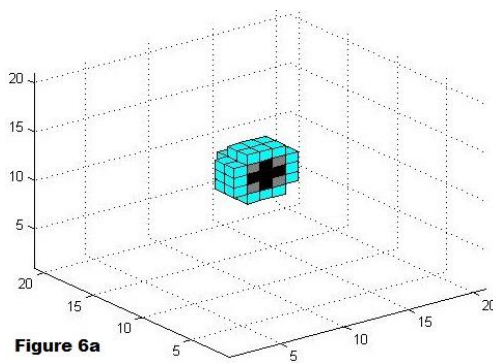 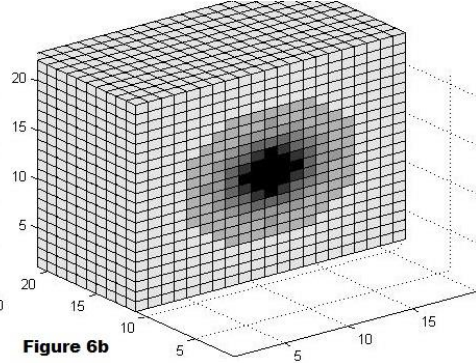 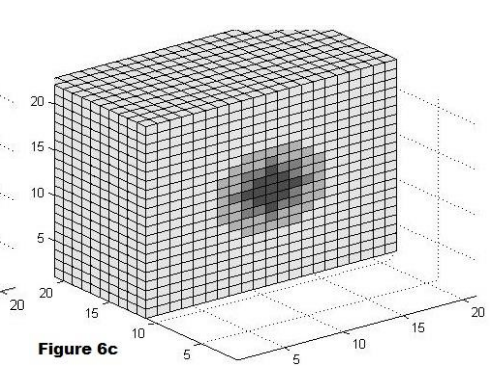

Figures 1-6 depict vertical sections of the graphical output of a simulation with parameter values as described in table 1, and $\lambda_1=5, \lambda_2=1.7$, $a_{max} = 16 \cdot 10^{-6}$ mitoses /10 sec. Total simulation time is 60 days, and the pictures are taken every 10 days. Figures 1a to 6a depict tumor expansion. Deep blue corresponds to voxels with total cell population above $C/2$ and more live than necrotic cells. Grey corresponds to voxels with total cell population above $C/2$ and more necrotic than live cells. Black corresponds to voxels with total cell population above $C/2$ and necrotic cell population above 1.5 times the living cell population within the voxel. Cyan corresponds to voxels containing total cell population less than $C/2$. Inspection of figures 1a to 6b show an agreement of the model with qualitative descriptions of tumor growth; Cells near the center of the tumor become necrotic, while an outer proliferating rim sustains tumor growth [29,51]. Figures 1b to 6b depict the spatiotemporal evolution of oxygen quantities in the lattice. Lightest color means that oxygen quantity within the voxel is intact. Black means that oxygen quantity within the voxel is below 15% of the respective oxygen quantity in intact tissue. The same holds for figures 1c to 6c, which depict the spatiotemporal evolution of glucose quantities. Lighter colors correspond to voxels with intact glucose levels

and darker colors to lower glucose levels (black corresponds to 40% of intact tissue). Apparently, despite the fact that cancer cells consume more glucose than oxygen, the impact of tumor growth in the oxygen levels of neighboring tissue is much more severe than that in glucose levels. This is due to the fact that glucose levels are much higher than that of oxygen and therefore less sensitive to the respective increased consumption required by the cancer cells. Interestingly, inspection of figures 3b,4b and 5b show that a severely hypoxic environment has been formed in the center of the tumor in the 30$^{th}$ day (fig. 3b), which is appears not so severe on the 40$^{th}$ day ( fig. 4b); This is due to the fact that once oxygen levels fall below $h_o F_o$ , cells are assumed to start becoming necrotic and seize consuming more oxygen than provided (Section V case 2), thereby allowing diffusion of oxygen from neighboring voxels to restore oxygen levels. This necrosis is depicted in figure 3b by the black voxels. However, this restoration is only temporary (fig 5b.) As soon as oxygen levels allow some proliferation, cells start consuming more oxygen resources than supplied, and once again oxygen levels fall below $h_o F_o$.

All simulations showed that $\lambda_2$ i.e. the parameter determining the oxygen consumption of proliferating cancer cells has the most critical impact on tumor growth. Figures 7a to 7d depict tumor growth curves for $a_{max} = 16 \cdot 10^{-6}$ mitoses / 10 sec, $\lambda_1$=2, and $\lambda_2$=3,4,5,6 respectively.

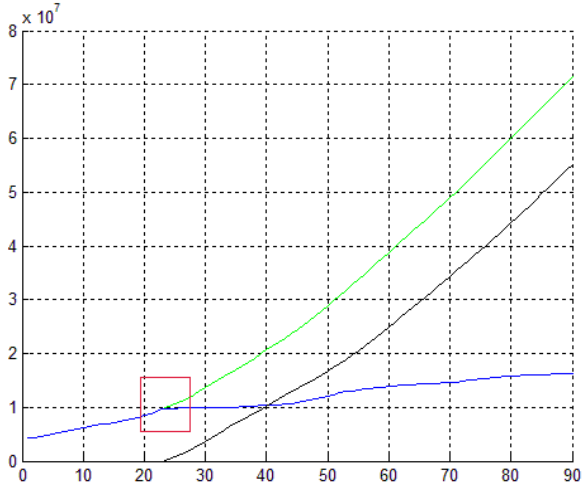

Figure 7a. Proliferating cancer cells consume 3 times more oxygen than quiescent host cells.

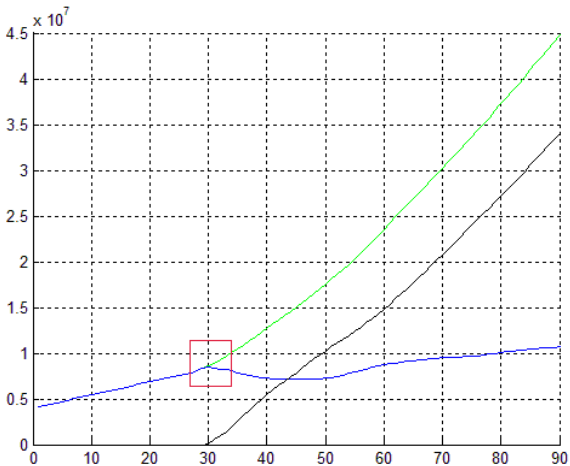

Figure 7b. Proliferating cancer cells consume 4 times more oxygen than quiescent host cells.

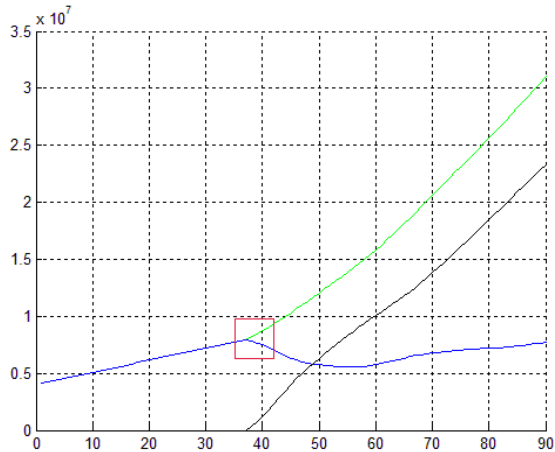
Figure 7c. Proliferating cancer cells consume 5 times more oxygen than quiescent host cells.

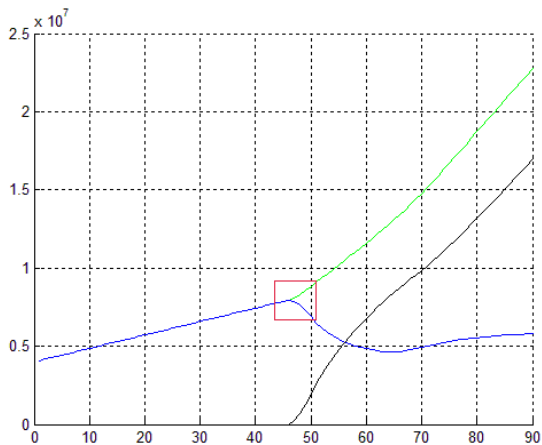
Figure 7d. Proliferating cancer cells consume 6 times more oxygen than quiescent host cells.
Figure 7. Blue line: living cells Black line: necrotic cells Green line: total (live+necrotic) cells.

In each case, the growth curve for living cells is initially exponentially increasing; All tumors reach a critical size, denoted by the red squares. Higher oxygen consumption implies lower (although not significantly) respective critical size and lower increase rate until this size is attended. Attending respective critical sizes is concurrent with the appearance of necrosis, and all cases show a decrease in the living cell populations, whose severity increases with higher oxygen consumption; This is expected, since higher oxygen consumption causes more severely hypoxic regions, where, after a certain point oxygen restoration cannot be performed at an adequate rate by oxygen diffusion from the neighboring cells. In all cases, living cell populations continue to grow at a linear rate, which decreases with increasing oxygen consumption.

Subsequently, we simulated the following scenarios. For maximum mitosis rates of $8 \cdot 10^{-6}$, $16 \cdot 10^{-6}$ mitoses/10 sec, corresponding respectively to doubling times of 10 and 5 days in ideal chemical conditions, we simulated growth curves for $\lambda_2 = 1.5$ and 2, and $\lambda_1 = 2, 4$ and 6.

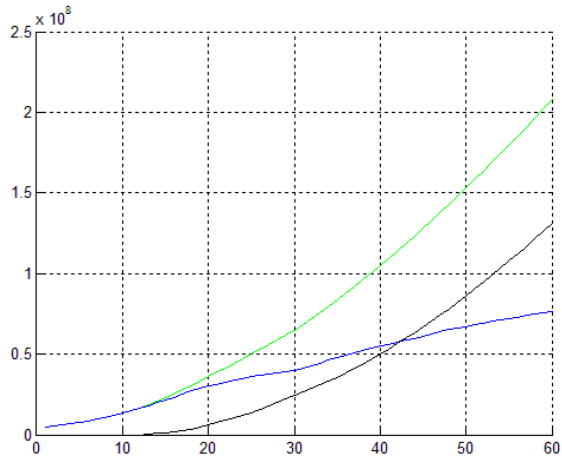

Figure 8a. Proliferating cancer cells consume 1.5 times more oxygen and 2 times more glucose than host quiescent cells.

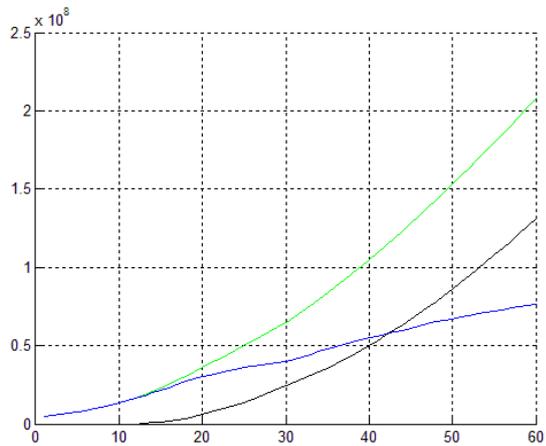

Figure 8b. . Proliferating cancer cells consume 1.5 times more oxygen and 4 times more glucose than host quiescent cells.

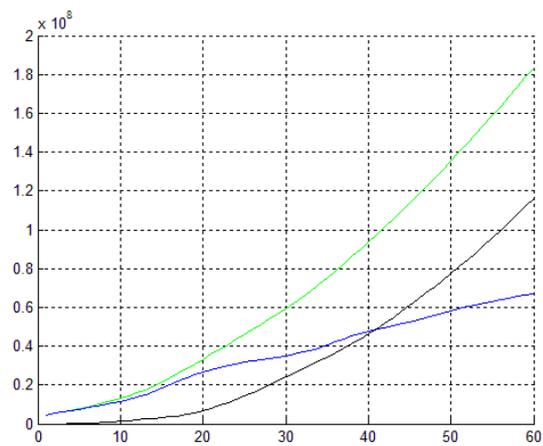

Figure 8c. . Proliferating cancer cells consume 1.5 times more oxygen and 6 times more glucose than host quiescent cells.

Figure 8. Blue line: living cells Black line: necrotic cells Green line: total (live+necrotic) cells. Doubling time in ideal conditions is 5 days.

In all figures 8a-8c, population of living cells shows an initial exponential increase, which later turns to linear. Glucose consumption appears to have an effect on tumor growth only in the third case (Figure 8c), where both necrosis and stopping of exponential growth appear earlier than in the other two cases, thereby affecting overall tumor growth.

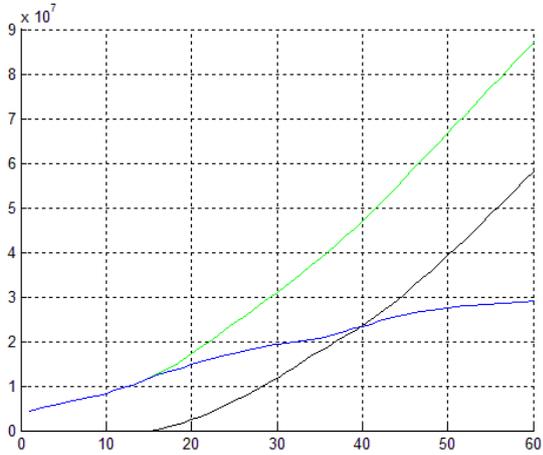

Figure 9a. Proliferating cancer cells consume 2 times more oxygen and 2 times more glucose than host quiescent cells.

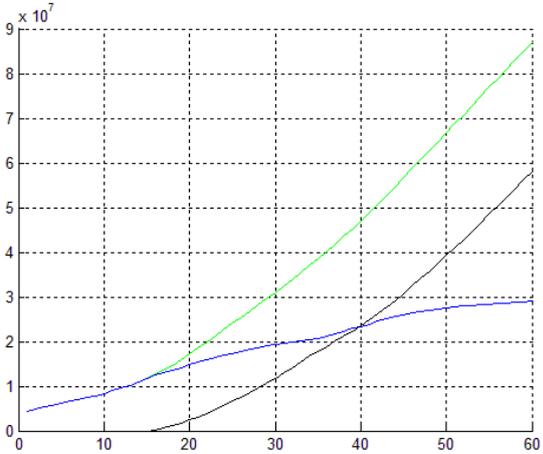

Figure 9b. Proliferating cancer cells consume 2 times more oxygen and 4 times more glucose than host quiescent cells.

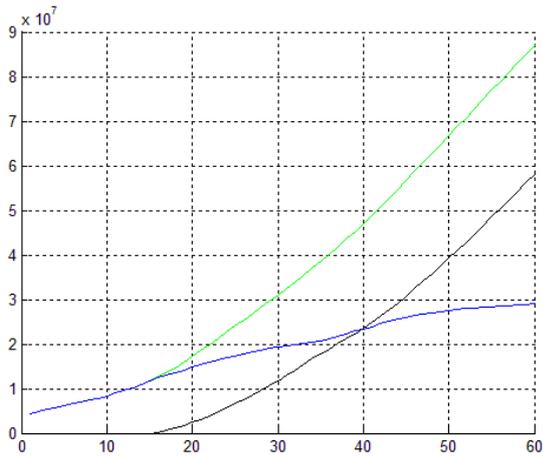
Figure 9c

Figure 9c. Proliferating cancer cells consume 2 times more oxygen and 6 times more glucose than host quiescent cells.
Figure 9. Blue line: living cells Black line: necrotic cells Green line: total (live+necrotic) cells. Doubling time in ideal conditions is 5 days.

In figures 9a-9c no impact of glucose consumption in overall tumor growth curves is observed. However, comparison of figures 9a-9a with figures 8a-8c reveals the impact of increased oxygen consumption. An increase in the order of 0.5 in $\lambda_2$ results in halving of the living cell population at day 60.

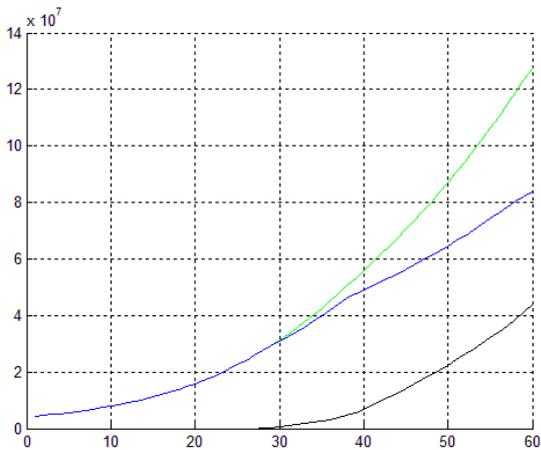
Figure 10a

Figure 10a. Proliferating cancer cells consume 1.5 times more oxygen and 2 times more glucose than host quiescent cells.

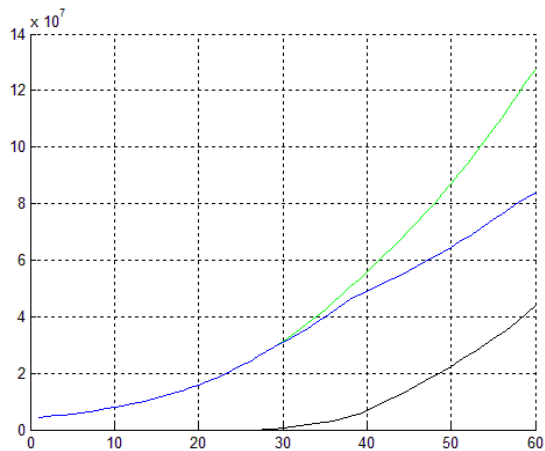

Figure 10b. Proliferating cancer cells consume 1.5 times more oxygen and 4 times more glucose than host quiescent cells.

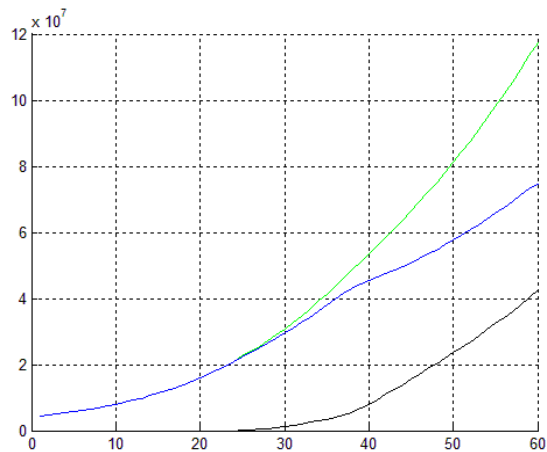

Figure 10c. Proliferating cancer cells consume 1.5 times more oxygen and 6 times more glucose than host quiescent cells.

Figure 10. Blue line: living cells Black line: necrotic cells Green line: total (live+necrotic) cells. Doubling time in ideal conditions is 10 days.

Comparison of figures 10a-10c with figures 8a-8c shows the effect of the maximum doubling time of the cancer cells, which is attained in ideal conditions. Increasing that parameter leads to slower tumor growth. However, during the entire simulation time total cell population consists mainly of live cells; necrosis starts later, necrosis rates are slower than those in figures 8a-8c, and necrotic population never exceeds living cells. As in figures 8a-8c, the impact of increased glucose consumption is detectable only in figure 10c, but is much slighter that in the case of figure 8c.

In figures 11a-11c necrosis starts later than in the respective cases depicted in figures 9a-9c and necrosis rates are much slower; The initial growth rate for living cells decreases slower than in the cases in figures 9a-9c, and necrotic population meets the living population at the end of the simulation time. However, comparison of figures 11a-11c with figures 10a-10c reveals the impact of increasing oxygen consumption; In all cases at the end of the simulation time the tumor is smaller. As in figures 9a-9c, there is no detectable impact of increased glucose consumption.

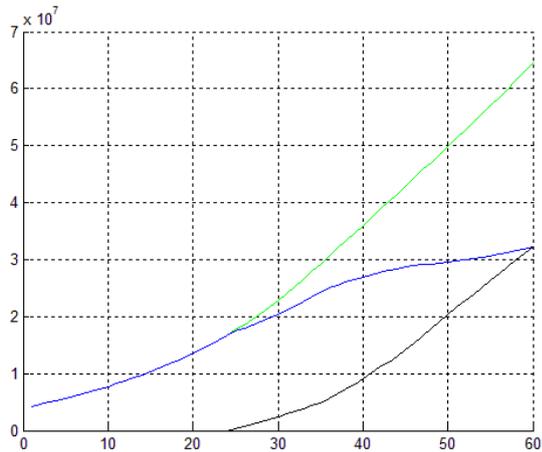

Figure 11a. Proliferating cancer cells consume 2 times more oxygen and 2 times more glucose than host quiescent cells.

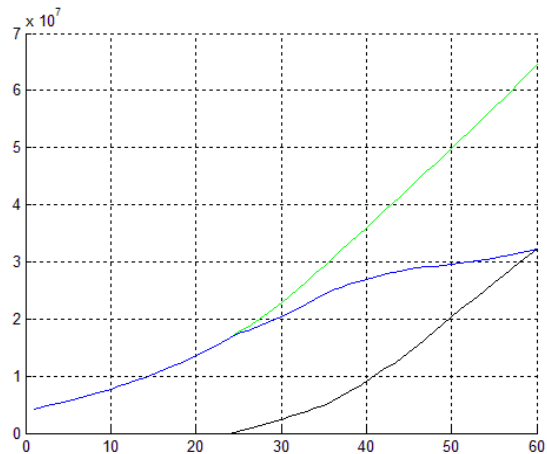

Figure 11b. Proliferating cancer cells consume 2 times more oxygen and 4 times more glucose than host quiescent cells.

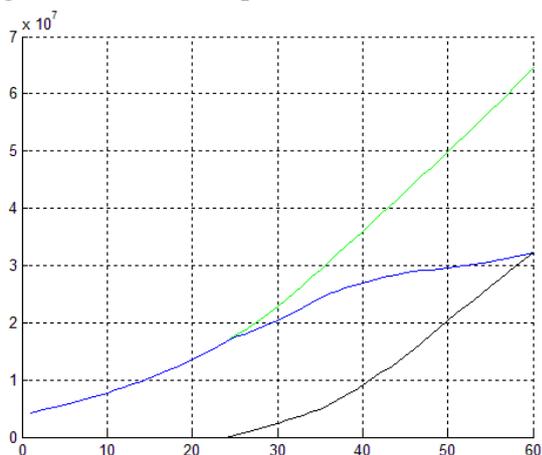

Figure 11c. Proliferating cancer cells consume 2 times more oxygen and 6 times more glucose than host quiescent cells.

Figure 11. Blue line: living cells Black line: necrotic cells Green line: total (live+necrotic) cells. Doubling time in ideal conditions is 10 days.

VI. Discussion and conclusions

We presented a modular framework for simulating tumor growth in a time varying chemical environment. Within this framework, we developed a model taking into account several phenomena observed during tumor progression, including diffusion of cells and molecules, consumption of chemical species by tumor cells, and cancer cell proliferation and necrosis. Simulations showed that for the scenarios tested, increased oxygen requirement by tumor cells can have a severe impact on tumor progression. In fact, the simulations depicted in figures 7a-7d show that when cells consume oxygen at higher rates than glucose (which phenomenologically corresponds to decreased utilization of glycolysis) the tumor practically cannot grow beyond a certain critical size; to grow any further, tumors must either decrease oxygen consumption and rely more on glucose (Figures 8,9,10,11) or establish their own blood supply network, thereby obtaining more oxygen resources. Angiogenesis is often associated associated with early responses to hypoxia [52,53]. Simulations showed a nontrivial interplay between oxygen requirements and maximum mitosis rates. In the avascular phase, cells affording a higher mitosis rate in ideal conditions do not necessarily work in favor of tumor growth. Faster proliferation requires increased oxygen consumptions and more tumor areas become severely hypoxic more quickly and more permanently; oxygen diffusion from neighboring voxel does not suffice to restore oxygen levels and consequently, necrosis rates are higher.

Computationally, a one-month simulation of tumor growth required 2.5 mins. For two-month simulations, this time rose to 5 or 6 mins. However, we note that many calculations performed by the model are parallelizable. For example, application of operators $T_2$ and $T_3$ to the state vector $(l_p, n_p, gl_p, o_p)$ could be performed independently. Furthermore, as previously mentioned algorithm 2 (or algorithm 6 in the appendix) are built on a (spatially distributed) dynamical systems perspective; They are also, eligible for parallelization.

Future work should consider more details concerning cell metabolism, including production/consumption of lactate, hydrogen ion production and the effects of extracellular pH [19,21,23]. Currently, a further development of the model is elaborated, where effects of neovascularization on local concentrations of nutrients are considered by introducing random fluctuations on the supply of these quantities in neighboring voxels of the tumor. The parallelizability potential of certain parts of the model should also be explored. Introducing anisotropy to cancer cell movement seems to be an interesting endeavor; The methods described in [38] provide means for Monte Carlo approximations of the probabilities presented in section III.

From a clinical perspective, any adaptation/validation scenario dictates that the model should be able to "bridge" the gap between clinical data taken at two distinct time points, with no surgery in between. In this context, extending the model with additional modules facilitating utilization of actual, clinically obtained imaging data and taking account of the effects of various treatment strategies like chemo- and radiotherapy is of major importance.

Appendix

Here we give implementation details on how to impose Neumann conditions on the boundary of the lattice. We will use the notation established in sections 1 and 3. Let $l_p$, $n_p$ denote the $N^3 \times 1$ vectors, whose entries are the populations of live and necrotic cells within each voxel at time instant $t_p$. Let $u_p$ denote the sum of $l_p$ and $n_p$, i.e. the total (live+necrotic) cell population within each voxel. For each voxel, let $F$ denote a common face neighbor, $E$ a common edge neighbor and $V$ a common face neighbor. Let $\Pr(A \to A), \Pr(A \to F), \Pr(A \to E), \Pr(A \to V)$ denote the probabilities calculated from equations (8) and (10) for voxels not lying at the boundary nor adjacent to it.

First, we construct two $N^3 \times 26$ matrices, denoted by $Neighbors$ and $Coefficients$ by the following algorithm

Algorithm 3

$Neighbors = N^3 \times 26$ zero matrix         //initialization
$Coefficients = N^3 \times 26$ zero matrix
**for** each triad $(i, j, k)$  $i, j, k = 1, ..., N$
  $counter = 0$
  **for** $di = -1, 0, 1$  $dj = -1, 0, -1$  $dk = -1, 0, 1$
    **if** the voxel with coordinates $(i + di, j + dj, k + dk)$ is in the lattice
      **if** $|di| + |dj| + |dk| = 1$         //common face neighbor
        $counter = counter + 1$
        $Neighbors(L(i, j, k), counter) = L(i + di, j + dj, k + dk)$
        $Coefficients(L(i, j, k), counter) = \Pr(A \to F)$
      **elseif** $|di| + |dj| + |dk| = 2$         //common edge neighbor
        $counter = counter + 1$
        $Neighbors(L(i, j, k), counter) = L(i + di, j + dj, k + dk)$
        $Coefficients(L(i, j, k), counter) = \Pr(A \to E)$
      **elseif** $|di| + |dj| + |dk| = 3$         //common vertice neighbor
        $counter = counter + 1$
        $Neighbors(L(i, j, k), counter) = L(i + di, j + dj, k + dk)$
        $Coefficients(L(i, j, k), counter) = \Pr(A \to V)$
      **end**
    **end**
  **end**
**end**

In view on the numbering of the voxels defined by the mapping $L$, we denote the $i$th voxel of the lattice by $A$. The $i$th row of the matrix $Neighbors$ contains the neighbors of $A$, as they are enumerated by $L$. Note that if $A$ lies at the boundary, the respective row has less than 26 nonzero entries. The nonzero entries of such a row can be 17, 11 or 7, depending on the position of the $A$. The entry at the $i$th row, $j$th column of the matrix $Neighbors$ is a neighbor $B$ of $A$, provided it is nonzero. The respective $(i, j)$ entry of the matrix $Coefficients$ is the number $\Pr(A \to B)$.

Apparently, the probabilities contained in rows with less than 26 nonzero entries are unnormalized. The following algorithm normalizes them. Initially, let $St$ denote a $N^3 \times 1$ vector, initially containing the number $\Pr(A \to A)$ in all of its entries.

Algorithm 4
**for** $i = 1$ to $N^3$
  **if** nonzero elements in $i$th row of the matrix $Neighbors$ are less than 26
    Calculate $sum = \sum_{j=1}^{26} Coefficents(i, j) + St(i)$
  $St(i) = St(i)/sum$
    **for** each $j$ such that $Neighbors(i, j)$ is nonzero
      $Coefficents(i, j) = Coefficents(i, j)/sum$
    **end**
  **end**
**end**

The previous algorithm normalizes the probabilities contained in each row of the matrix $Coefficents$ and the vector $St$, and at the same time. The final step is to construct a $N^3 \times 26$ matrix denoted $Probabilities$ with the following property: Using the notation of the previous paragraph, the $i$th row-$j$th column entry of that matrix should be the probability $\Pr(B \to A)$. This matrix, along with the vector $St$ will subsequently be used to algorithmically implement the modeling of living cell diffusion. The matrix $Probabilities$ is constructed as follows:

Algorithm 5

$Probabilities = N^3 \times 26$ zero matrix                      //initialization
**for** $i=1$ to $N^3$
   **for** each $j = 1$ to 26 such that $Neighbors(i,j)$ is nonzero
     Set $k = Neighbors(i,j)$
     Find the $m$th column of $Neighbors$ such that $Neighbors(k,m) = i$
     Set $Probabilities(i,j) = Coefficients(k,m)$
   **end**
**end**

Finally, the algorithmic implementation of living cell diffusion model with Neumann boundary conditions is as follows.

Algorithm 6

**for** each voxel $A$ in the lattice with resp. coordinates $(i,j,k)$
   **if** voxel $A$ and each of its neighbors have zero total population
     $next\_l(A) = l(A)$             // no diffusion
   e**lse**
     **if** $u(A) \geq C$
       $s = St(L(i,j,k))l(A)$
     **else**
       $s = l(A)$
     **end**
     **for** each $m = 1$ to 26 such that $Neighbors(L(i,j,k), m)$ is nonzero
        **if** $u(B_i) \geq C$
         $s = s + Probabilities(L(i,j,k), m)l(B_i)$
        **end**
     **end**
     $next\_l(A) = s$
   **end**
**end**
$l = next\_l$

Again, this algorithm leaves the total population of living cells unchanged. We note that the biological meaning reflected on algorithm 6 is slightly different from that on algorithm 2 in the main text. In algorithm 6, for the living cells within any voxel to start invading neighboring voxels, cancer cells within it should have first reached a critical total (live+necrotic) population. In algorithm 2, for any movement of cells between two neighboring voxels to happen, the total number of cancer cells (live +necrotic) should have reached a critical population in at least one of the two voxels. However, both algorithms are based on the same core ideas. From the performed simulations, it seems that there are no major differences in the results. To keep the presentation of the core ideas as simple as possible, we opted to put the first version in the main text.

Competing interests
The authors have no competing interests of any kind.

Funding
This work has been supported in part by the European Commission under the project Computational Horizons in Cancer (CHIC): Developing Meta- and Hyper-Multiscale Models and Repositories for In Silico Oncology (FP7-ICT-2011–9, Grant agreement no: 600841), MyHealthAvatar: ADemonstration of 4D Digital Avatar Infrastructure for Access of Complete Patient Information (FP7-ICT-2011–9–600929), and p-Medicine: Personalized Medicine (FP7-ICT−2009.5.3–270089). The authors confirm that the funder had no influence over the study design and content of the article.

Author Contributions
MA conceived the mathematical ideas and developed the respective arguments and algorithms. GS coordinated the work and proposed the scenarios for the provision of clinical input to the model via multiple tomographic data. MA developed the simulation codes and produced the results. MA wrote the first draft of the manuscript. MA and GS revised, corrected, read and approved the manuscript.